\documentclass[12pt,preprint,letterpaper,xdvi]{aastex}
\usepackage{amsmath} 
\usepackage{wrapfig} 
\usepackage{array}  
\usepackage{natbib}
\usepackage{brief_bib}
\usepackage{lscape}
\usepackage{verbatim}
\usepackage{rotating}
		
\newcommand\bvec{{\bf B}}
\newcommand\br{{\bf B}_{\mathrm{ref}}}
\newcommand\bp{{\bf B}_{\mathrm{P}}}
\newcommand\pot{_{\mathrm{P}}}
\newcommand\jvec{{\bf J}}

\newcommand\avec{{\bf A}}
\newcommand\nhat{{\bf \hat{n}}}

\newcommand\rvec{{\bf r}}

\newcommand\al{$\alpha$}
\newcommand\als{$\alpha\mbox{ }$}

\newcommand{\vol}{{\cal V}}
\newcommand{\path}{{\cal L}}

\begin{document}

\title{Guiding nonlinear Force-Free Modeling Using Coronal Observations: First Results Using a Quasi Grad-Rubin scheme}
\author{A.~Malanushenko$^{1,2}$, C.~J.~Schrijver$^2$, M.~L.~DeRosa$^2$, M.~S.~Wheatland$^3$, S.~A.~Gilchrist$^3$}
\affil{$^1$Department of Physics, Montana State University, Bozeman, MT, USA\\
       $^2$Lockheed Martin Advanced Technology Center, Palo Alto, CA, USA\\
       $^3$Sydney Institute for Astronomy, School of Physics, University of Sydney, Australia}

\begin{abstract}
Presently, many models of the coronal magnetic field rely on photospheric vector magnetograms but these data have been shown to be problematic as the sole boundary information for nonlinear force-free field (NLFFF) extrapolations. Magnetic fields in the corona manifest themselves in high-energy images (X-rays and EUV) in the shapes of coronal loops, providing an additional constraint that at present is not used due to the mathematical complications of incorporating such input into numerical models. Projection effects and the limited number of usable loops further complicate the use of coronal information. We develop and test an algorithm to use images showing coronal loops in the modeling of the solar coronal magnetic field. We first fit projected field lines with field lines of constant-\als force-free fields to approximate the three-dimensional distribution of currents in the corona along a sparse set of trajectories. We then apply a Grad-Rubin-like iterative technique to obtain a volume-filling nonlinear force-free model of the magnetic field, modifying method presented in \citet{Wheatland2007}.  We thoroughly test the technique on known analytical and solar-like model  magnetic fields previously used for comparing different extrapolation techniques  \citep{Schrijver2006, Schrijver2008} and compare the results with those obtained  by presently available methods that rely only on the photospheric data.  We conclude that we have developed a functioning method of modeling the coronal magnetic field by combining the line-of-sight component of photospheric magnetic field with information from coronal images. Vector magnetograms over the full or partial photospheric boundary of the numerical domain could optionally be used. 
\end{abstract}

\section{Introduction}\label{sec_intro}

The ability to build adequate models of the coronal magnetic field is extremely important for understanding the physics of the solar corona. The corona is believed to be generally in a force-free (or at least low-$\beta$) state \citep{Gary2001}. Destabilization of this state may lead to eruptions, with contributing factors including topological properties of the field, such as the existence of null points and excessive magnetic twist \citep{Canfield1999}. The amount of energy released in eruptions cannot exceed the amount of free magnetic energy at the time of destabilization. Moreover, as the coronal field generally evolves in such a way that its total helicity only changes due to helicity flux across the photosphere and into the heliosphere \citep{Berger1984}, the assessment of helicity at one point in time, such as prior to a CME, might be beneficial for studies of the evolution of the corona and heliosphere. Modeling of coronal heating is frequently performed as 1-D hydrodynamic (or static) models following magnetic field lines using values of magnetic field along these field lines as important input \citep[e.g.,][]{Lundquist2008}. So this modeling would also benefit from better models for the magnetic field.

The general problem of constructing a force-free magnetic field (hereafter FFF) to model the coronal field is formulated as follows \citep{Nakagawa1971}. The objective is to find a magnetic field $\bvec$ which satisfies the divergence-free condition  
\begin{equation}
\nabla\cdot\bvec=0
\label{div_free}
\end{equation}
and the force-free equation 
\begin{equation}
\nabla\times\bvec=\alpha\bvec, 
\label{fff}
\end{equation}
where $\alpha$ is a proportionality constant between the magnetic field and magnetic current\footnote{The parameter \als has topological meaning associated with the amount of twist in the field, e.g., \citet{Gold1960}.}. Equations~(\ref{div_free}) and~(\ref{fff}) must be solved for $\bvec$ and \als in a volume domain $\vol$ subject to boundary conditions $\bvec|_{\partial\vol}$ (or $\bvec\cdot\nhat|_{\partial\vol}$ and $\alpha|_{\partial\vol}$). The problem is not in general linear and the solution is hence called a \textit{``nonlinear force-free field''}, hereafter NLFFF. Particular cases include a \textit{linear} force-free field (hereafter LFFF) that solves the system assuming $\alpha(\rvec)=\mbox{const}$, or a potential field (where $\alpha(\rvec)=0$) which we refer to in the text as $\bp$. 

Many difficulties arise when solving the problem of constructing a non-linear force-free field. The underlying reasons for these difficulties are physical, mathematical and computational. Physically, the full vector magnetic field at the lower boundary $z=0$ is presently obtained only in the photosphere, where plasma forces are significant. That is to say, Equation~(\ref{fff}) is not appropriate at the lower boundary level \citep{Gary2001}. Also, the component of $\bvec$ transverse to the line of sight at the photosphere is subject to an intrinsic 180$^\circ$ ambiguity, and measurements of boundary data at the top and side boundaries of the computational domain are not available at this time \citep[see][for an extensive discussion of these issues]{Demoulin1997b}. Typically, assumptions are made about the side boundaries, e.g., a field matching a potential source surface model \citep{Schrijver2003} is assumed, and there are various methods to resolve the azimuthal ambiguity \citep{Metcalf2006}. Mathematically, the system is nonlinear and at the present stage the uniqueness, and even the existence, of a solution in general for a given boundary conditions are not proven. Finally, there are computational difficulties that have to do with the high instrumental uncertainty in the measurements of the transverse horizontal component of the photospheric magnetic field and the small spatial scale of current changes, possibly below the instrumental resolution, in the lower boundary. This uncertainty has more impact than it might seem at first sight because $\bvec\cdot\nabla\alpha=0$, implying $\alpha=\mbox{const}$ along magnetic field lines\footnote{This follows from Equation~(\ref{fff}) by taking the divergence of both sides of the equation.}. Hence, field lines must connect points with the same \als on positive and negative polarities at the lower boundary so the boundaries must have equal amounts of incoming and outgoing magnetic flux for each value of \al. Noise in \als at the lower boundary and limits to the field of view prevent this condition from being satisfied, and the problem is in general ill-posed. Techniques exist for ``pre-processing'' of the boundary data to attempt to mitigate this problem \citep[e.g.,][]{Wiegelmann2006, Wiegelmann2008}. 

The existing methods to address the difficulties outlined above do not appear to be developed to a level such that photospheric vector magnetograms may be used to reliably model the coronal field. Different methods for solving the NLFFF problem, and even different implementations of the same method, applied to the same photospheric data, and even the same method applied to different polarities of the same data, frequently yield results inconsistent with each other and with the coronal features \citep{Schrijver2006, Metcalf2008, Schrijver2008, DeRosa2009}. Such methods are, for example, the magnetofrictional relaxation \citep[e.g., ][]{Ballegooijen2004}, optimization \citep[e.g., ][]{Wiegelmann2004} and the Grad-Rubin method \citep[e.g., ][]{Wheatland2007}. \\

Extensive studies are needed to address all of these issues. Hence, a substantial time might pass before reliable vector magnetograms consistent with the upper chromosphere become available for models of the coronal field. Until they are available another source of information is needed for modeling the coronal fields. We propose this source to be coronal loops. 

Coronal loops, observed in X-Ray and EUV images, are believed to follow lines of the magnetic field, and therefore they should be of help for magnetic extrapolations. Unlike vector magnetograms, this information originates in the force-free corona, where Equation~(\ref{fff}) is appropriate. Field lines spread apart with height and so do bundles of coronal loops \citep[though the field generally expands with height, individual loops are found to have nearly constant diameter with height, see][]{Klimchuk2000}. Consequently, the structure of the magnetic field in the corona should be less fine than at the photospheric level so it might in principle be better resolved by currently available instruments. Observed loops also give an idea about the overall connectivity of the coronal field, which might otherwise be easily distorted by even minor noise present in photospheric vector magnetograms and therefore in \al, as discussed above. 

Even if techniques of processing vector magnetograms are developed to the point that NLFFF models are generally reliable, coronal loops as an additional constraint might be of great benefit, for example, for studies of energy release in solar flares. Vector magnetograms undergo relatively minor changes during even major flares \citep[e.g., ][found only a fractional change in the transverse component of the field in a small patch of the active region during a large X-class flare]{Wang2012}. On the contrary, the changes in the connectivity of the coronal magnetic field can be large-scale and dramatic even in smaller flares. As the connectivity of the magnetic field manifests itself in the shapes of coronal loops, the latter provide a powerful guide for tracking sudden changes in the field. 

Making use of coronal loops is, however, a non-trivial task. The plasma is optically thin, and what is observed by instruments is the integrated emission of all the plasma along the line of sight. Extracting individual loops from bundles of overlapping loops is a non-trivial image processing task, with the possible exception of isolated loops far away from the core of the regions. Some progress, though, has been made in this direction  \citep[e.g.,][]{Aschwanden2008}. Another difficulty is that all currently existing instruments, with the exception of STEREO satellites \citep{StereoRef}, only observe the Sun in one projection, so the three-dimensional structure of the loops is not immediately obvious. 

Recently, substantial progress has been made in studies of coronal loops as magnetic features. \citet{Lim2007} first fitted observed projections of coronal loops with lines of a LFFF. \citet{Malanushenko2009b} developed a semi-automatic algorithm for such fits, applicable to portions of loops, and showed that \als values obtained this way statistically correlate with \als values for a NLFFF model. Progress also has been made in obtaining information from the original images. Numerous studies, \citep[e.g.,][]{Aschwanden2009} have demonstrated good results on triangulating loops trajectories using STEREO data. 

Two approaches to the use of image data in magnetic modeling are as follows. Reconstructed 3-D loop trajectories and \als values along them may be determined approximately using the scheme from \citet{Malanushenko2009b}, hereafter the MLM09 fit. This provides information at least about the 3-D trajectories of some field lines and \als in the corona along these field lines. Stereoscopically-derived data offers another possibility: the inferred 3-D loop trajectories could be used in conjunction with values of the vector magnetic field at the loop foot points. Vector magnetograms are of course prone to the problems outlined above. However, in the case of using loops, the field values only need to be accessed \textit{at a sparse set of locations} in the lower boundary. If it is possible to estimate the chromospheric magnetic field (assuming, for example, that the field does not change much with height in the chromosphere) in \textit{at least a few} patches in an active region, and provided that these patches contain foot points of the stereoscopically determined loops trajectories, then this information could be used as in the first approach, but with more accurate results. 

In this paper we propose a new method of constructing a NLFFF using such information derived from coronal loops. We also draw attention to the value of coronal loop observations for magnetic modeling in general. Such methods might in principle be of use in areas of plasma physics other than coronal studies. It might for example be desirable in laboratory plasma studies to estimate what kind of a force-free field would have a required topology and magnitude of currents. 

The paper is organized as follows. In Section~\ref{sec_method} we describe the quasi Grad-Rubin scheme enabling us to make use of coronal loops with and without vector magnetograms. In Section~\ref{sec_input_data} we discuss various inputs. Section~\ref{sec_tests} describes the general scheme of a set of tests of the method and figures of merit obtained. The results of the tests are presented in detail in Section~\ref{sec_appl}. Section~\ref{sec_summ} discusses the results, evaluating how successful the scheme is and its value for modeling of the coronal field.

\clearpage
\section{Description Of The Quasi Grad-Rubin Method}\label{sec_method}

Suppose there is a domain $\vol$ with boundary $\partial\vol$ and the following are given: 
\begin{enumerate}
	\item[(a)]{$\bvec\cdot\nhat|_{\partial\vol}$ (where $\nhat$ is the normal to $\partial\vol$);}
	\item[(b)]{a set of trajectories $\{\path_i\}_{i=1}^{N}$ in $\vol$ along which the force-free parameter values $\{\alpha_i\}_{i=1}^{N}$ are known (and are constant along each individual trajectory).}
\end{enumerate}
The objective is to find the field $\bvec$ that solves Equation~(\ref{fff}) and matches the boundary conditions (a) and the volume constraints (b).

The procedure is iterative and is similar to a Grad-Rubin iteration \citep{Grad1958}. It starts with potential field $\bvec^{(0)}=\bvec_{P}$ as an initial guess for the field and an initial guess $\alpha^{(0)}$ for the force-free parameter, which at each point in the domain is set equal to $\alpha_i$ for the closest point in the volume for which an $\alpha_i$ value is known. Then on every $n$-th iteration the updated cubes $\bvec^{(n)}$ and $\alpha^{(n)}$ are obtained as follows:
\begin{enumerate}
	\item{Impose the volume constraints by setting $\alpha^{(n-1)}=\alpha_i$ along the trajectories $\path_i$.}
	\item{Calculate updated field values $\bvec^{(n)}$ from  \begin{equation}\nabla\times\bvec^{(n)}=\alpha^{(n-1)}\bvec^{(n-1)}\label{curl_iter}\end{equation} subject to the prescribed boundary conditions. This equation is solved using a vector potential $\avec^{(n)}$ such that $\bvec^{(n)}=\nabla\times\avec^{(n)}$, so the divergence-free condition is satisfied to truncation error.}
  \item{Calculate an updated set of values for the force-free parameter $\alpha^{(n)}$: for every point in $\vol$, assign $\alpha^{(n)}=\langle\alpha^{(n-1)}\rangle$ \textit{averaged along the field line in $\bvec^{(n)}$ that passes through that point}. If a field line leaves the domain through any boundary but the lower one, the value of $\alpha$ is set to zero along it (in common with the Wheatland~2007 GR scheme). This ensures that no currents go off to infinity so that the fields' energy remains finite.}
	\item{Repeat 1.-3. until $\bvec^{(n)}\approx\bvec^{(n-1)}$ and $\alpha^{(n)}\approx\alpha^{(n-1)}$ to within a tolerance and therefore $\nabla\times\bvec^{(n)}\approx\alpha^{(n)}\bvec^{(n)}$.}
\end{enumerate}

This sequence is similar to an existing Grad-Rubin method of solution of the NLFFF problem \citep{Wheatland2007, Wheatland2009}. The only difference between these two schemes is how the updated $\alpha^{(n)}$ cube is calculated in the Step~3. In both schemes, at each point in the domain a field line is traced in $\bvec^{(n)}$. In the original Grad-Rubin code, positive or negative polarity is picked at the lower boundary; and $\alpha^{(n)}$ at each point in the volume is set to the \textit{value at the boundary point} in the chosen polarity where it is crossed by the field line. The only exception is the boundary itself for all points in the chosen polarity where \als is kept constant. In the Quasi Grad-Rubin, $\alpha^{(n)}$ at each point is assigned the \textit{average} of \als from the \textit{previous iteration} along this field line. The only exception is the volume constraint paths where \als keeps the volume constraint value.

Let us consider some particular simple cases. First of all, it is clear that if the initial guess for $\bvec$ and \als already satisfies $\nabla\times\bvec=\alpha\bvec$, then the scheme keeps it unchanged. If the field or $\alpha$ differ from a solution to Equation~(\ref{fff}) even at one point, the field changes -- though if it is only one point that is different, convergence is achieved in a single iteration. It is also clear that if $\alpha=0$ everywhere, currents do not appear; since the scheme averages $\alpha$ at every iteration it is incapable of introducing $|\alpha|\geq \max(|\alpha_i|)$. These cases make sense: if the answer is close to the correct answer, convergence is achieved rapidly, and if no currents are specified to start with, the scheme does not change the input potential field. 

But what happens in an intermediate situation: currents are known on \textit{some}, but not all flux tubes in the domain? Can a solution be reached at all? If several solutions are plausible given the constraints, which solution (of any) is achieved? There are proofs of existence and uniqueness of solution for the force-free problem, achievable by Grad-Rubin iteration, if \als is sufficiently small in some sense \citep[see][]{Bineau1972}. However, the range of \als is not clearly defined and it is unclear whether solar-like fields are within this range. If they are outside of this range, that does not mean, the original Grad-Rubin iteration necessarily fails. It is unclear if similar proofs exist for the proposed Quasi Grad-Rubin scheme. In the absence of this, we consider a process of numerical experimentation to test the scheme. We attempt also to determine how many field line trajectories $\path_i$ are sufficient to enable convergence. In Section~\ref{sec_appl} we review some of our experiments.

\section{Different Types of the Input Data for Quasi Grad-Rubin Scheme}\label{sec_input_data}

The Quasi Grad-Rubin numerical scheme (hereafter ``QGR'', in contrast to ``GR'' for Grad-Rubin algorithm) could in principle be used with \als constrained at \textit{any} set of locations including the lower boundary. Hence it may be used with vector magnetograms, setting ${\alpha_i}$ at the lower boundary to the vector magnetogram derived value, 
\begin{equation}\alpha|_{z=0}=\left.\frac{1}{B_z}\left(\frac{\partial B_y}{\partial x} -\frac{\partial B_x}{\partial y}\right)\right|_{z=0}.\end{equation}
Vector magnetograms may also be used with or without the loop trajectories. We identify three different kinds of inputs for QGR: \als along loop trajectories; \als along loop trajectories and at the lower boundary; and \als at the lower boundary only\footnote{Note that \als does \textit{not} have to be constrained at all points on the lower boundary.}. Traditional schemes are designed to work for the last case only. 

If the \als values, wherever set, are approximate, this will introduce uncertainties. To properly test QGR we try to recover several known fields and use both approximated trajectories and those drawn from the known field. We note that there may be applications of QGR even if it would not work with approximate data, for example for problems of the following kind: constructing a magnetic field with twist prescribed along certain trajectories. It could also be used with stereoscopically triangulated loops and currents derived from the photospheric vector magnetograms, or \als values and trajectories obtained by other means. 

In this paper we test various possible inputs and compare the results with the reference fields. Table~\ref{input_data_types} outlines the degrees of freedom available for such tests. We refer to various inputs and schemes using this chart, e.g., II.b is QGR applied to volume constraints alone, drawn from the reference field. 

\begin{table}[!hc]
 \caption{\small{Possible combinations of different inputs to QGR, in addition to $\bvec\cdot\nhat|_{\partial\vol}$. Note that I.a has the same input as the original GR scheme but the algorithm is different, so QGR with I.a input is \textit{not} equivalent to GR.}} 
 \centering 
 \begin{tabular}{m{11.0cm}m{2.0cm}m{2.0cm}}
  & & \\
  \hline
  & \multicolumn{2}{c}{Values of \als at $z=0$} \\
  Values of \als along loop trajectories & Known & Unknown \\
  \hline
  None & I.a & --- \\
  From the field lines of the model field (``ideal'' input) & I.b & II.b \\
  From the MLM09 approximation derived from 2-D projections of these field lines (realistic coronal input) & I.c & II.c \\
  \hline
  & & \\
 \end{tabular}
 \label{input_data_types}
\end{table}

The QGR scheme was implemented by modifying an existing GR code (which we refer to as CFit version 1.3), described in \citet{Wheatland2009}. It is a ``self-consistent'' scheme, in that it picks a polarity, finds a force-free solution using boundary data from that polarity, hence obtaining values of \als everywhere in the volume including the other polarity at the boundary; then the \als map at the other polarity is updated with the weighted average of the values obtained from this new solution and the values that existed before this solution was found. It then repeats the cycle using the updated \als map from the other polarity. The cycles are continued until the two solutions obtained using \als values at the opposite polarities are consistent with each other to a tolerance. Since QGR does not use the lower boundary in the same way, we discard the switching between cycles and modify the way the \als values in the volume are calculated (see previous section). Also, instead of using $B_z$ at $z=0$ and a 2-D array of $\alpha$ values at $z=0$ the modified code uses $B_z$ at $z=0$ and two 3-D arrays: $\alpha_i$ along the trajectories and an initial guess everywhere else and a ``mask'', i.e. another 3-D array with entries either unity along ${\path_i}$ or zero everywhere else. These are the two principal changes to the CFit code. The calculation of the field on a given iteration, i.e. the solution of $\nabla\times\bvec^{(n+1)}=\alpha^{(n)}\bvec^{(n)}$ is unchanged. The code uses a vector potential for this step and hence the field satisfies $\nabla\cdot\bvec^{(n+1)}=0$. 

\section{Description Of Metrics for QGR Solutions}\label{sec_tests}

In this paper we try to recover several known force-free fields, namely those from \citet{Schrijver2006} and \citet{Schrijver2008}. The iterations are  initialized with the same potential fields used in the original studies. We also try to construct a NLFFF based on a dipole magnetogram and two specified loop trajectories (with a reference field which is not known). 

We estimate the quality of the reconstruction and the relative force-freeness of the known solutions using metrics of which most have become standard in NLFFF modeling. These are as follows. 

\begin{itemize}
	\item{$E/E\pot$ versus $E_\mathrm{ref}/E\pot$: energy in the reconstructed field versus energy of the reference field (for a perfect reconstruction the two would be equal).}
	\item{$H(\bvec|\bp)$ versus $H(\br|\bp)$: relative helicity\footnote{By $H(\bvec_1|\bvec_2)$ we mean the helicity of the field $\bvec_1$ relative to the field $\bvec_2$.} in the reconstructed field versus that of the reference field (for a perfect reconstruction the two would be equal).}
	\item{$\mbox{CWsin}=(\sum{|\sin\theta||\jvec|})/(\sum{|\jvec|})$ versus $\mbox{CWsin}_{\mathrm{ref}}$, where $|\sin\theta|=|\jvec\times\bvec|/|\jvec||\bvec|$: the total current-weighted sine of the angle between $\bvec$ and $\jvec$ (for a perfectly force-free field this is zero).}
	\item{Metrics of similarity between $\bvec$ and $\br$, normalized to equal unity if $\bvec=\br$:
				\begin{itemize}
					\item{$\mbox{C}_{\mathrm{CS}}= \frac{1}{N}\sum{[\bvec\cdot\br/(|\bvec||\br|)]}$ (where $N$ is the number of points in the domain): the average cosine of the angle between $\bvec$ and $\br$;} 
					\item{$\mbox{C}_{\mathrm{vec}}= (\sum{\bvec\cdot\br})/(\sum{|\bvec||\br|})$: same as previous but with increased weight in regions of stronger field;} 
					\item{$E_m'=1-E_m$, where $E_m=\frac{1}{N}\sum{|\bvec-\br|/|\br|}$: the average relative difference between $\bvec$ and $\br$;} 
					\item{$E_n'=1-E_n$, where $E_n=\sum{|\bvec-\br|}/\sum{|\br|}$: same as previous but with increased weight in regions of stronger field.} 
				\end{itemize}
				}
\end{itemize}

We omit metrics for how well $\nabla\cdot\bvec$ is satisfied, because the method \citep[in common with ][]{Wheatland2009}, uses a vector potential to calculate the field and hence achieves a divergence-free state to truncation error \citep{Press1992}.

\section{Sample Applications of QGR}\label{sec_appl}
\subsection{QGR Solution for a Dipole Field}\label{sec_dipole} 
The first test case is a simple dipole field aligned in the E-W direction with the North half of both magnetic poles having negative twist and the South half of both magnetic poles having matching positive twist. This model could be viewed as a simple representation of an emerged untwisted flux rope whose foot points became distorted in such a way that the field at one (leading) polarity has been inclined more than the field at the second (following) one, perhaps due to subsurface flows. Such a difference in inclinations is observed for solar active regions \citep{Howard1991_7}. 

\begin{figure}[!hc]
 \begin{center}
  \includegraphics{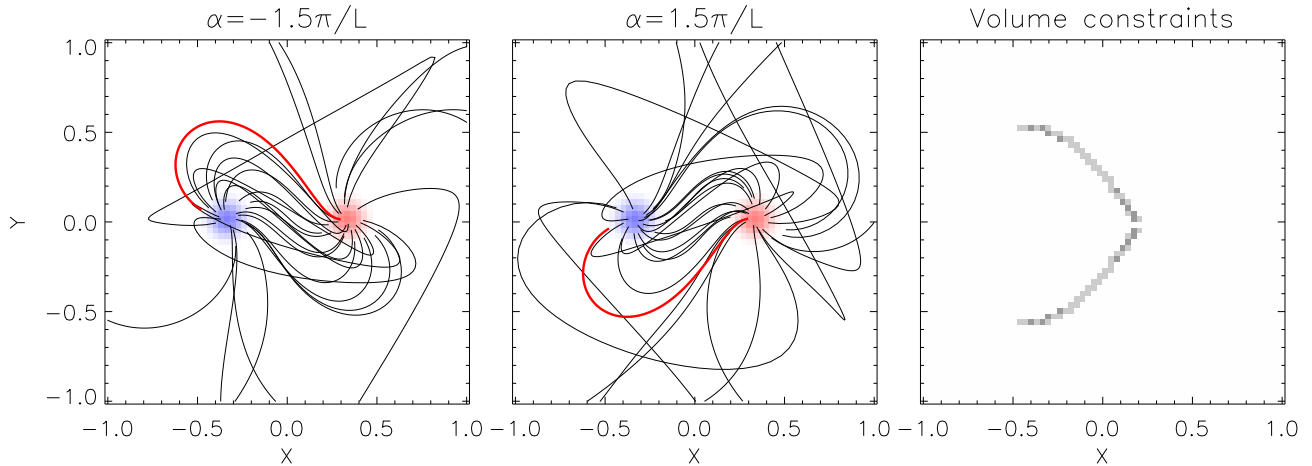}
  \end{center}
 \caption{Input data for the dipole field test case: two field lines (thick red) drawn from two constant-\als fields (left and middle panels). The right panel shows points where the volume constraints are applied (i.e., the points through which the trajectories $\path_i$ pass). Points in the north and south halves of the domain are assigned $\alpha_i$ values of $\pm\alpha_0$, respectively.}
 \label{dipole_input}
 \end{figure}

To construct the field we calculate two constant-\als fields confined to half spaces \citep{Chiu1977} with equal and opposite twist ($\alpha_0=\pm1.5\pi/L$, where $L$ is the size of the domain). We draw one field line for $\path_i$ from each of these and use these field lines (which imitate coronal loops) as volume constraints. The initial guess for \als is $\pm\alpha_0$ in the S and N halves, respectively. The fields, field lines, and the locations of the constraints are shown in Figure~\ref{dipole_input}.

 \begin{figure}[!hc]
 \begin{center}
  \includegraphics{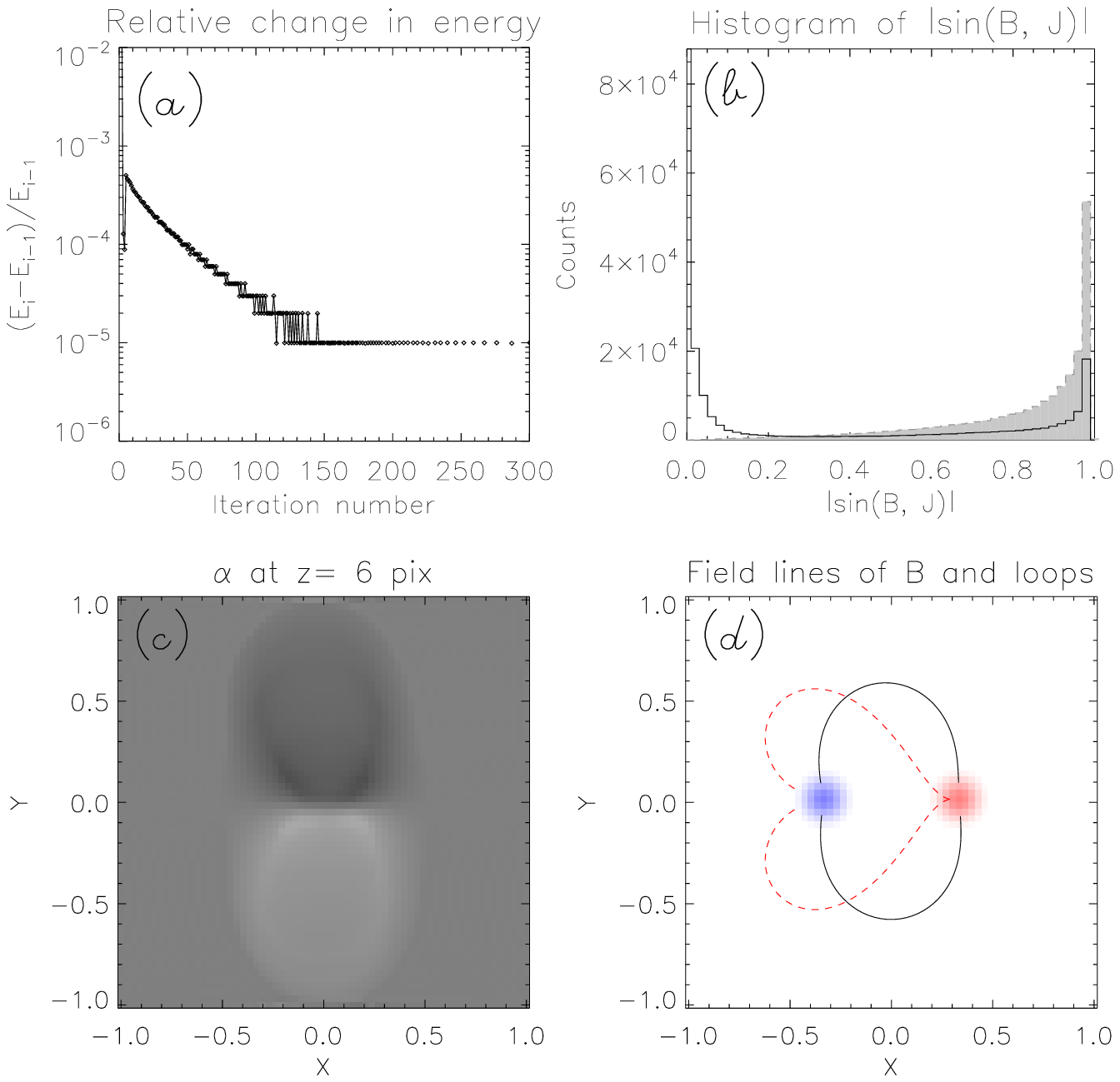}
  \end{center}
 \caption{The solution for the dipole test case. The QGR iteration converges as shown in panel (a), which displays the change in the free energy of the field consecutive iterations. The constructed field $\bvec$ is force-free, as shown in panel (b) by a histogram of $|\sin(\bvec, \jvec)|$ for $\bp$ (shaded curve) and $\bvec$ (solid curve). The peak at unity is due to the current-free regions in the field, as explained in the previous section. The initial field $\bp$ is current-free and $\bvec$ retains current-free regions (in particular where field lines leave the domain through side or top boundary). The field $\bvec$ is nonlinear, as seen in panel (c), which shows a horizontal slice of \als at a height of six pixels in the box (out of 64). In this panel black corresponds to negative \als and white to positive \al. The field lines of $\bvec$, however, do not match the constraints! Field lines of $\bvec$ are shown in panel (d) as solid black while the constraints are dashed red lines. Field lines are initiated at midpoints, rather than foot points, of the loops.}
 \label{dipole_underest}
 \end{figure}
 
QGR is found to converge to a nonlinear force-free field, as shown in Figure~\ref{dipole_underest}. Achievement of the force-free state is shown by the distribution of $\sin(\bvec, \jvec)$ peaking at zero. A second smaller peak at one is a contribution from the current-free regions of the field, and is due to $\jvec$ in these regions being exclusively numeric noise. To support this, the shaded histogram on Figure~\ref{dipole_underest} shows the distribution of the same quantity evaluated for potential field. The nonlinearity of the field, that is, the presence of different \als values on different field lines, is illustrated by a distribution of \als on a horizontal slice. These additional diagrams are shown in Figure~\ref{dipole_underest}~(a)-(c).

The field lines of the solution, interestingly, are found to follow different trajectories than the two constraining field lines. In fact, the solution, having \als close to the two imposed constraints in the two halves, appears much closer to a potential field than the highly twisted constant-\als fields we drew the loops from! This is shown in Figure~\ref{dipole_underest}~(d) which illustrates the constraining loops and two field lines of the solution initiated at the midpoints of these loops. 

\clearpage

The fact that the NLFFF solution with currents similar in magnitude as in the \citet{Chiu1977} constant-\als field appears to have much less twisted field lines is due to at least two effects. First, as open field lines are required to carry no currents in the NLFFF model constructed by a QGR \citep[in common with the][implementation of GR]{Wheatland2007}, the volume containing currents has to be contained in the computational domain, while the \citet{Chiu1977} fields have currents in the entire half space. As the volume containing currents in the NLFFF is smaller, the current density has to be stronger for the field lines to have similar shape. Secondly, currents which run in opposing directions close to each other might counteract each other in influencing the shape of field lines.

\begin{figure}[!hc]
 \begin{center}
  \includegraphics[height=22cm]{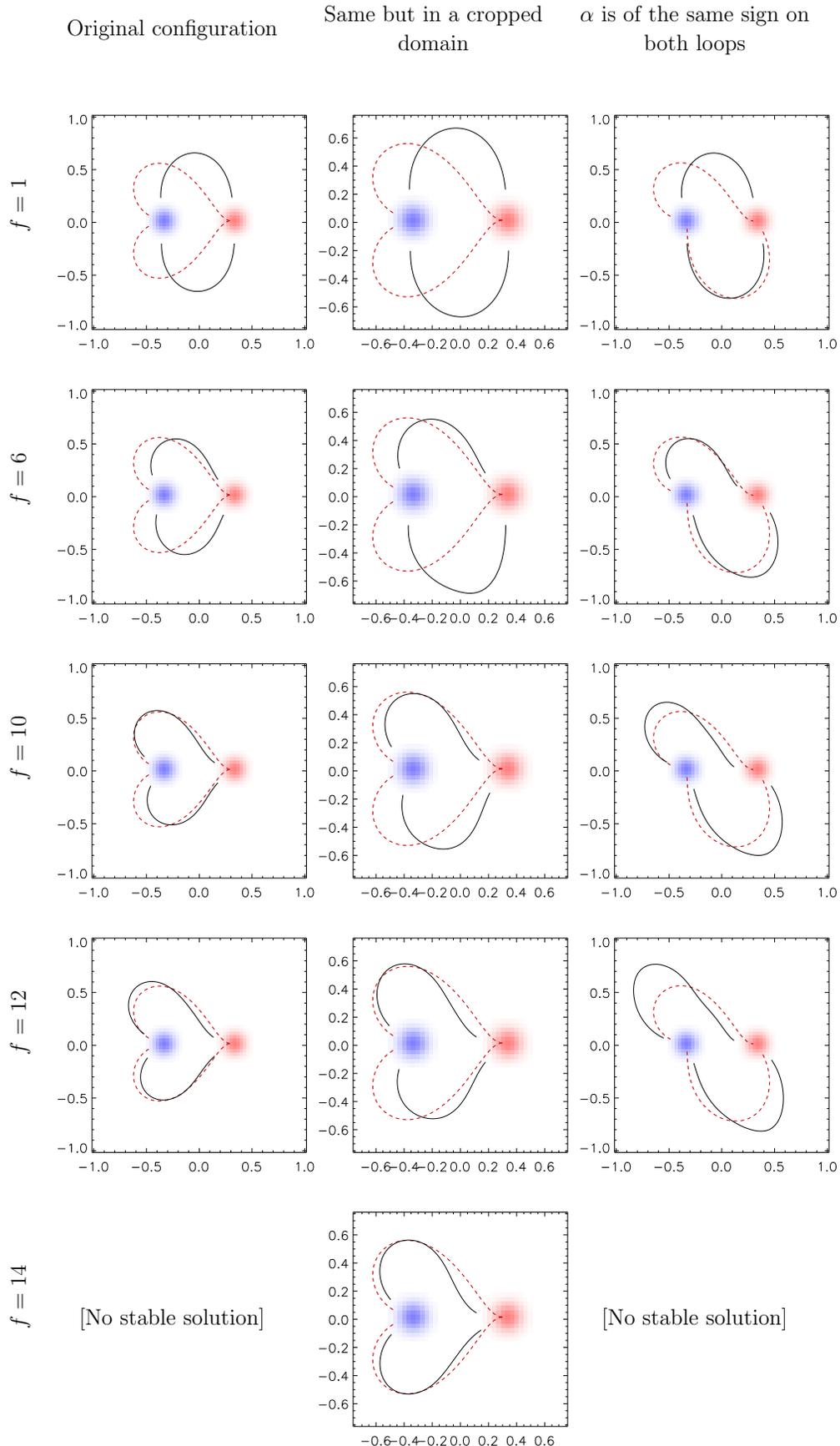} 
  \end{center}
 \caption{Experiments on the dipolar test case to illustrate the effect of opposing currents and domain size on the scaling factor $f$. See Section~\ref{sec_dipole} for description. The notation is the same as in Figure~\ref{dipole_underest}.}
 \label{dipole_factor}
 \end{figure}

To illustrate these effects we perform a simple numeric experiment. We repeat the computation but set $\alpha_i=\pm f\alpha_0$ on the constraining paths for several values of $f$. The results are shown in Figure~\ref{dipole_factor}, first column. We also repeat the experiment for a box of smaller size (cropped on the sides and the top) but with otherwise identical setup (Figure~\ref{dipole_factor}, second column). Finally we repeat the computation with the same lower boundary but different volume constraints: a field line in the lower half of the domain is drawn from a constant-\als field with the same sign of \als as the upper half but smaller in magnitude, with $\alpha=-\alpha_0/4=-3\pi/8 L$ (Figure~\ref{dipole_factor}, third column). For the original setup, the best match between the constraining loops (red dashed curves) and field lines of the solution (black solid curves) initiated at the mid points of these loops is achieved with $f\approx10-12$, while if the same (or at least similar) field lines are required to exist in a field with currents confined to a smaller domain, the best match between the loops and the solution is achieved with $f\approx14$. If both loops have \als of the same sign, the best match is achieved for $f\approx 6-10$. (No steady solution is found for $f=14$ for the setup in the first and third columns; in this case the QGR solution continues to oscillate. The solutions for $f=14$ in the second column and $f=12$ in the third columns exhibits oscillations but of small magnitude. Later we discuss these oscillations which may result from the input being inconsistent with a NLFFF solution and describe a procedure to damp them. In this section, however, our point is to illustrate the significance of the scale factor $f$.)

The scale factor $f$ cannot be evaluated \textit{a priori}, but it may be estimated using observables, i.e., coronal loops. This can be done by minimizing the difference between the shapes of coronal loops and field lines of different solutions corresponding to different values of $f$. In the next two sections we demonstrate that scaling factors obtained this way are indeed a proportionality constant between \als on lines from a NLFFF and \als of the approximation of these field lines by lines of constant-\als fields of the type constructed by \citet{Chiu1977}. 

\hoffset=-0.5cm
\voffset=-0.5cm

\clearpage
\subsection{QGR Solutions for Low \& Lou Fields}\label{sec_llf} 
In this section we try to reconstruct the test field from \citet{Schrijver2006} (the ``reference field'' further in this section) using QGR. This test case is a member of family of analytic NLFFF's introduced by \citet{LowLou1990}. 

For the first QGR test we use field lines of the reference field as trajectories $\path_i$ and take $\alpha_i$ values corresponding to the correct \als values for the reference field. Physically, this could correspond to stereoscopically derived loops with chromospheric vector field known around their foot points. These data are hard to obtain at present; we use them mainly to test QGR alone, that is, on the ``ideal'' data not contaminated by measurement errors. We use 113 randomly selected field lines $\path_{i,\mathrm{ref}}$ (out of a bigger sample, chosen to be closed field lines, i.e. with both foot points on the lower boundary). We calculate $\alpha_{i,\mathrm{ref}}$ numerically everywhere in $\br$ and evaluate it along these 113 field lines. We discard all but 27 of these, retaining the ones that may be well fitted using MLM09, to make this test consistent with the next, realistic test presented later in this section. We construct two solutions of the same size as the reference field ($64^3$ pixels), with and without the additional constraint of vector field data at the lower boundary (i.e., schemes I.b and II.b from Table~\ref{input_data_types}). Figures of merit are calculated in the same subdomain as in \citet{Schrijver2006}. This center subdomain is also used to estimate a best-matching scaling factor $f$.

\begin{figure}[!hc]
 \begin{center}
  \includegraphics{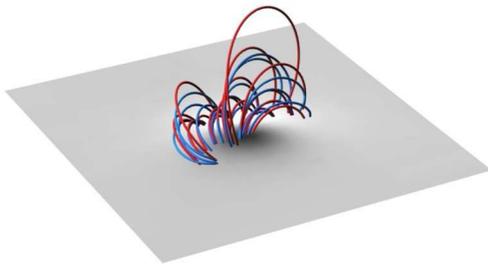} 
  \end{center}
 \caption{Results of MLM09 applied to the reference field from \citet{Schrijver2006}, used for I.c and II.c inputs as trajectories $\path_i$: $\br$ field lines (red) and MLM09 field lines (blue).}
 \label{loop_fit_llf}
 \end{figure}

In the second test we determine $\path_i$ and $\alpha_i$ based on a fit to the resulting MLM09 field derived from the normal field at the lower boundary as in \citet{Malanushenko2009b}. We calculate QGR solutions using these data with and without vector field data at the lower boundary (I.c and II.c from Table~\ref{input_data_types}). This tests the applicability of QGR to realistically available coronal data. It is not obvious that using approximations to loop trajectories and approximate values of \als along them is sufficient to create a field model at least as good as those derived from vector magnetogram data. To investigate this we project the same 113 field lines as for the ideal data to the $z=0$ plane to simulate the appearance of loops in the plane of the sky. We treat these 2-D projections as synthetic loops and use them to obtain \als values $\alpha_{i,\mathrm{MLM09}}$ along trajectories $\path_{i,\mathrm{MLM09}}$ using the MLM09 fit procedure. We discard loops which upon visual examination are poorly fitted, which leaves 27 loops that appear to have a near perfect fit. The trajectories of these loops and their MLM09 approximations are shown in Figure~\ref{loop_fit_llf}.

\begin{figure}[!hc]
 \begin{center}
  \includegraphics[height=12cm]{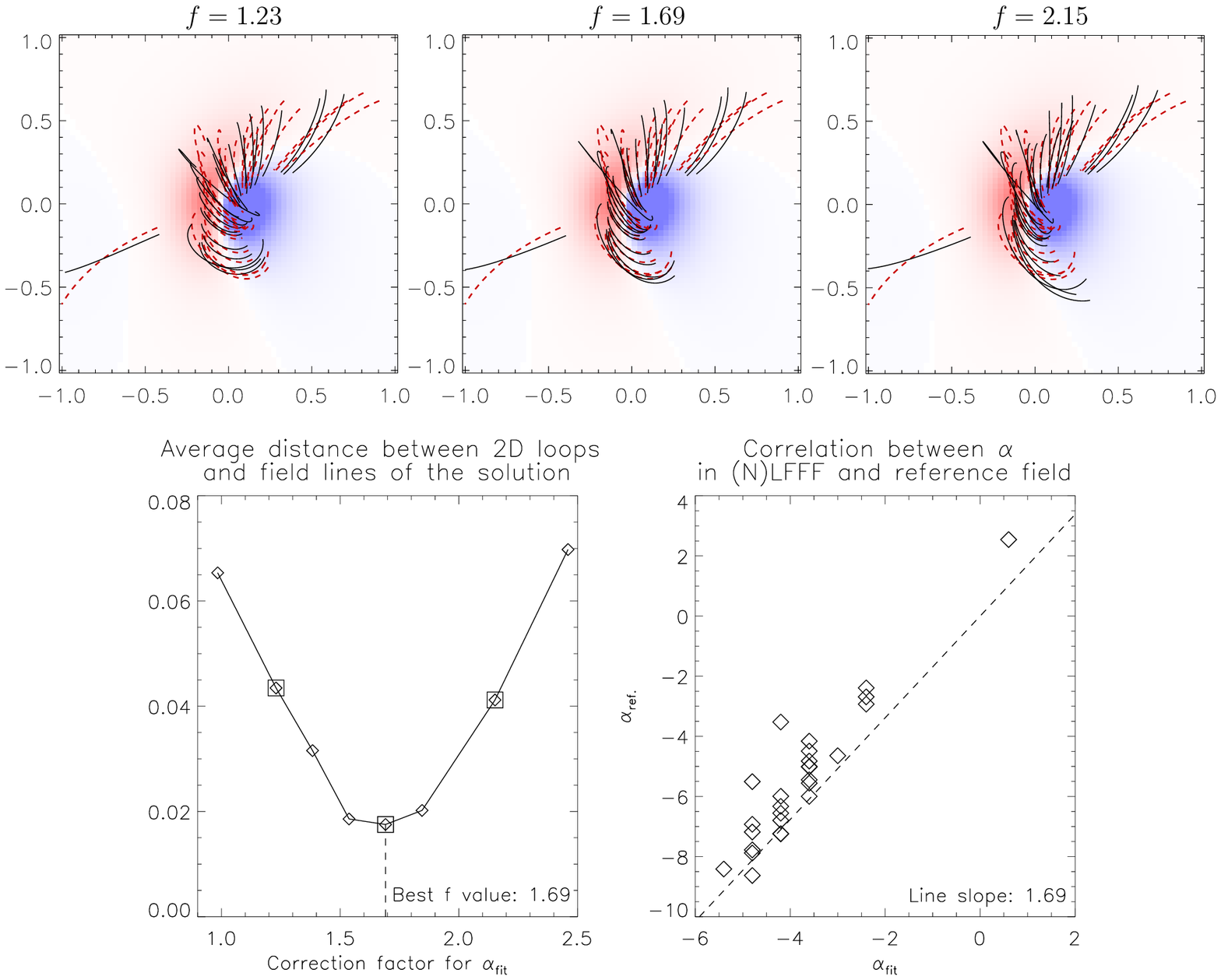} 
  \end{center}
 \caption{Top row: several QGR solutions for the reference field from \citet{Schrijver2006} for input II.c (see Table~\ref{input_data_types}) corresponding to the same input but different scaling factor $f$ for \al. Red dashed lines show lines of $\br$ projected onto $z=0$ and used as loops which were approximated by MLM09 to construct trajectories $\path_i$. Black lines show corresponding lines of $\bvec$ (initiated at midpoints of the lines of $\br$). Field lines in the core of the domain are the most affected by the choice of $f$. Bottom left: average distance between loops (projections of lines of $\br$) and corresponding lines of $\bvec$ in projection onto the plane of the sky ($z=0$). The difference is smallest for $f\approx1.69$. This coefficient is remarkably close to the scaling coefficient between $\alpha_{\mathrm{ref}}$ and $\alpha_{\mathrm{fit}}$ from MLM09, shown at the lower right panel as a dashed line. Diamonds show $\alpha_{\mathrm{ref}}$ of the individual field lines of $\br$ versus $\alpha_{\mathrm{fit}}$ of the MLM09 approximations of these field lines.}
 \label{fact_llf}
 \end{figure}

Keeping in mind the results from Section~\ref{sec_dipole}, we calculate several solutions for II.c with seven different scaling factors $f$ applied to input \als but with the calculations otherwise identical. Three of these solutions (corresponding to $f=1.23$, 1.69 and 2.15) are shown in Figure~\ref{fact_llf}, top row. For each of the solutions we estimate how closely the field lines match the synthetic 2-D loops used to construct the trajectories $\path_i$. We calculate the average distance between the projected loops and the corresponding field lines of the solution. The result (as a function of $f$) for all seven solutions is shown in Figure~\ref{fact_llf} (bottom left plot). The solution for $f\approx1.69$ is the closest match to the loops. The bottom right panel in the same figure is a scatter plot of $\alpha_{\mathrm{fit}}$ and $\alpha_{\mathrm{ref}}$ for individual loops. The fit appears to underestimate the value of \al. The underestimation factor is remarkably close to 1.69 (dashed line on the same plot). This suggests that such underestimation \textit{could be derived a posteriori} from the observed loops.

\begin{figure}[!hc]
  \begin{center}
    \includegraphics{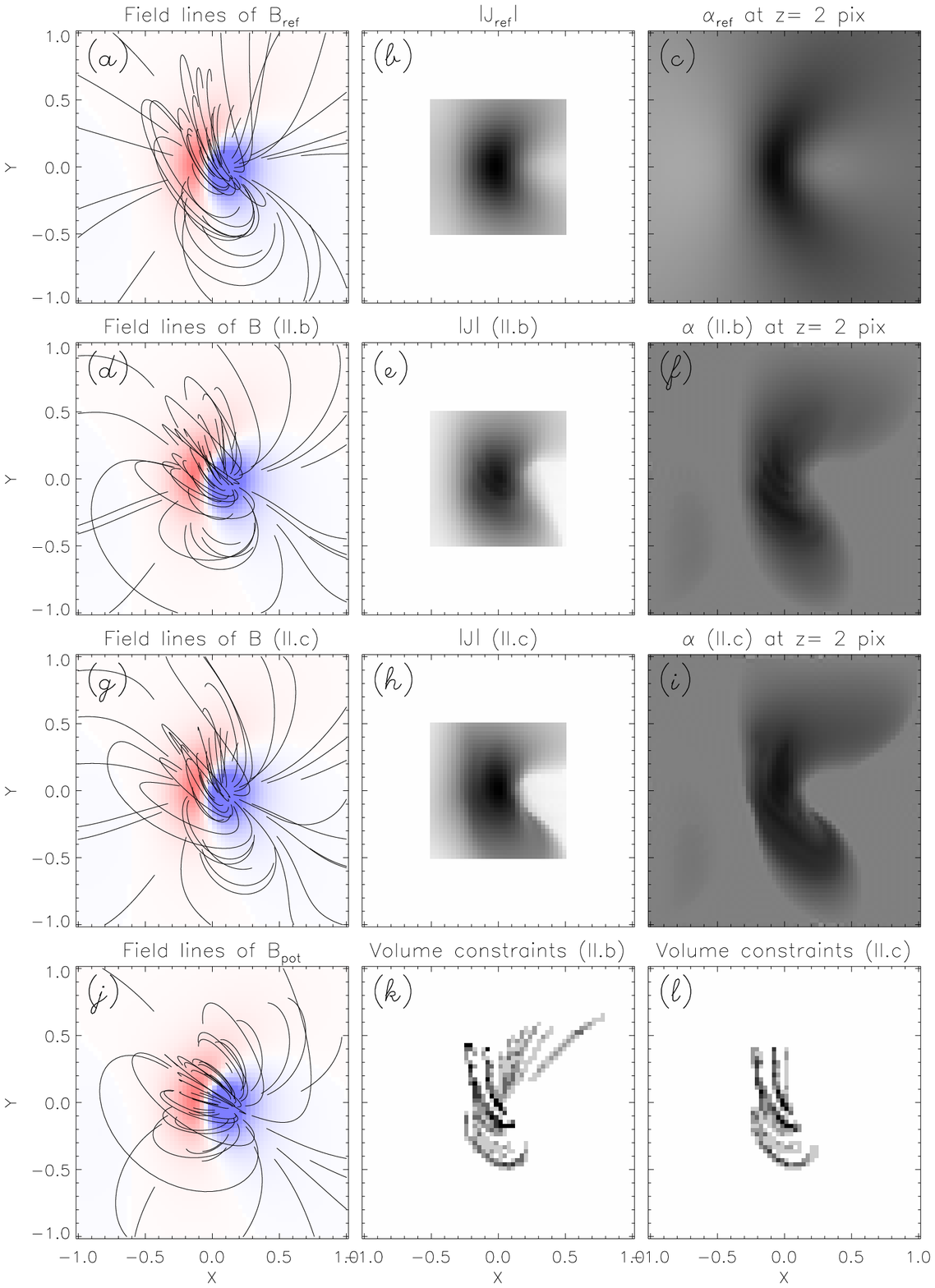} 
  \end{center}
  \caption{Reconstruction of Low \& Lou field from \citet{Schrijver2008} using schemes II.b and II.c, i.e. QGR with ideal and realistic loop input, that is, reconstructed from 2D loop projections (see Table~\ref{input_data_types}). Panels \textit{(a)}-\textit{(i)}: field lines, line-of-sight integrated magnitude of current and horizontal slices of \als for $\br$ and $\bvec$ for II.b and II.c. Panel \textit{(j)}: field lines of $\bvec_{P}$ (all field lines are traced from the same starting points). Panels \textit{(h)}, \textit{(l)}: line-of-sight integrated volume constraints for II.b and II.c.}
 \label{llf/small_table}
 \end{figure}

\begin{figure}[!hc]
  \begin{center}
   \includegraphics{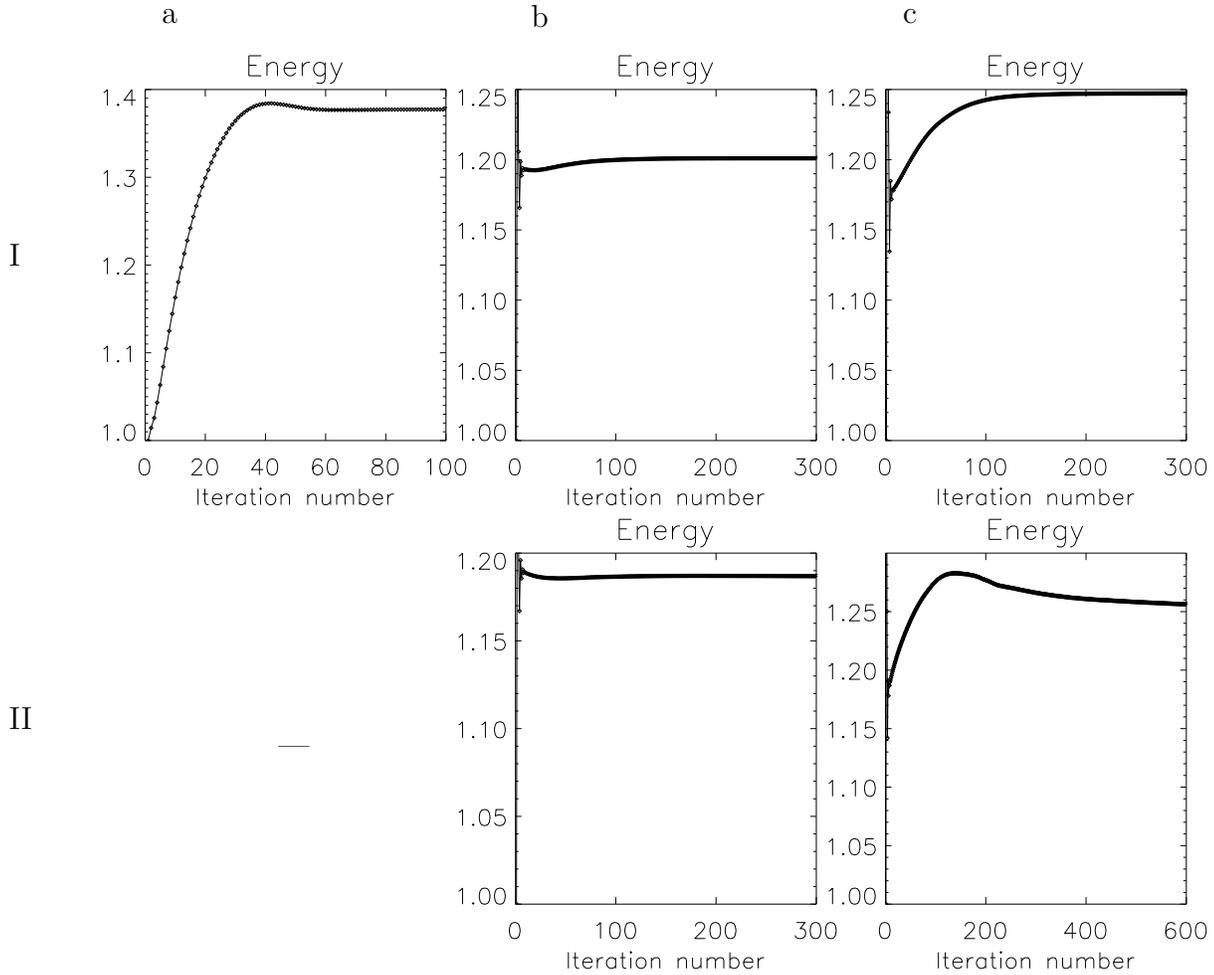} 
  \end{center}
  \caption{Energy $E/E_P$ at each iteration as demonstration of the convergence of the QGR iteration for the Low \& Lou field test case. The energy on these plots is shown for the entire domain \citep[while Table~\ref{table_llf} reports the numbers for the middle sub-domain identical to the one in][as do the other tables in the manuscript]{Schrijver2006}. Top row, from left to right: I.a, I.b and I.c solutions (refer to Table~\ref{input_data_types} for notation). Bottom row, from left to right: II.b, II.c.}
 \label{llf/energies}
 \end{figure}

The results of our tests (I.c and II.c with $f=1.69$) are summarized in Table~\ref{table_llf} and Figure~\ref{llf/small_table}. Figure~\ref{llf/energies} shows that convergence is achieved for all solutions. We conclude that QGR is able to reconstruct the reference field, including the shape of field lines\footnote{Excepting the field lines leaving the computational domain, which are required to carry no current in our scheme, but in fact do carry currents in the Low \& Lou field.}, the structure of the currents and the distribution of \al, remarkably well. The  figures of merit show that the QGR reconstructions are at least as close (and typically clos\textit{er}) to the right answer as other methods. In particular, the \textit{smallest} estimate for the free energy we obtain is \textit{closer} to the correct answer than any of the estimates based on the vector field boundary values alone reported by \citet{Schrijver2006}.  


\hoffset=-1cm
\begin{table}[!hc]
	\centering
\begin{tabular}{m{1.0cm}m{1.9cm}m{1.9cm}m{1.9cm}m{1.9cm}m{1.9cm}m{1.9cm}c}
 & $\mbox{C}_{\mathrm{vec}}$ & $\mbox{C}_{\mathrm{CS}}$ & $1-E_n$ & $1-E_m$ & CWsin & $E/E\pot$ & $H(\bvec|\bp)$ \\
& & & & & & & \\
\multicolumn{8}{l}{\textit{Reference field}}\\
& 1.00 & 1.00 & 1.00 & 1.00 & 0.01 & 1.24 & 1.00 \\
& & & & & & & \\
\multicolumn{8}{l}{\textit{Quasi Grad-Rubin with vector magnetograms}}\\
I.a & 0.99	& 0.93	& 0.80	& 0.64	& 0.03	& 1.27 & 0.67 \\
I.b & 0.99	& 0.96	& 0.83	& 0.70	& 0.02	& 1.19 & 0.81 \\
I.c & 1.00	& 0.97	& 0.88	& 0.75	& 0.02	& 1.23 & 0.91 \\
& & & & & & & \\
\multicolumn{8}{l}{\textit{Quasi Grad-Rubin with loop trajectories alone}}\\
II.b & 0.99	& 0.96	& 0.81	& 0.68	& 0.02	& 1.17 & 0.77 \\
II.c & 0.99	& 0.97	& 0.86	& 0.78	& 0.02	& 1.23 & 0.93 \\
& & & & & & & \\
\multicolumn{8}{l}{\textit{Ranges reported in \citet{Schrijver2006}}}\\
& 0.94 -- 1.00 & 0.54 -- 0.91 & 0.48 -- 0.92 & -2.2 -- 0.66 & 0.03 -- 0.57 & 0.82 -- 1.14 & --- \\
& & & & & & & \\
\multicolumn{8}{l}{\textit{Potential field}}\\
 & 0.86	& 0.87	& 0.50	& 0.44	& --- &	1.00 & 0.00 \\
\end{tabular}
\caption{Metrics (defined in Section~\ref{sec_tests}) for different QGR solutions with different types of input data applied to the \citet{Schrijver2006} test case. The values for I.c and II.c are reported for the optimal solution with $f=1.69$. The values for $\br$ and $\bp$ are shown for comparison, and so are the ranges of values for different NLFFF extrapolations reported in \citet{Schrijver2006}. Relative helicity is stated in fractions of that of the reference field. For notation, refer to Table~\ref{input_data_types}.}
\label{table_llf}
\end{table}

 \voffset=-1.5cm
 \hoffset=-1cm
 
 \voffset=-1.5cm
 \hoffset=-1cm
 
\clearpage 
\subsection{QGR Applied to a Solar-Like Field}\label{sec_karel}

In this section, we investigate whether QGR is applicable to solar data. The \citet{LowLou1990} family of fields has axial symmetry which is not in general observed in active regions, and both magnetic field and current vary unrealistically smoothly through the lower boundary by comparison with vector magnetogram data. Hence, we repeat the experiments from the previous section, but with choosing a more realistic solar-like field as $\br$. We use two particular NLFFF solutions from \citet{Schrijver2008}, who presented NLFFF reconstructions of the coronal field for AR 10930 before and after a major flare using several extrapolation methods applied to \textit{Hinode} vector magnetograms. They found that the extrapolations which best matched observed coronal features were GR solutions obtained with the \citet{Wheatland2007} code, using \als values from the positive polarity of the magnetograms (hereafter Wh$^{+}_{\mathrm{pp}}$). We use those as our reference field. These solutions also had the largest free energy of all extrapolations. Another advantage of these fields for our study is that they use the same boundary conditions and nearly the same numeric implementation as the QGR scheme. We emphasize that the objective is in this case \textit{not} to create a realistic representation of coronal field but to test the new algorithm on a \textit{known} NLFFFs that are expected to more closely resemble the coronal field overlying a solar active region. 

For both pre- and post-flare reference fields we select random sets of field lines and evaluate $\langle\alpha\rangle$ on each field line. These field lines are used as trajectories $\path_i$ in II.b set-up and their projections onto $z=0$ plane are used as loops for II.c set-up. We also use $B_z$ at the lower boundary and start with the same initial $\bp$ as the other schemes in \citet{Schrijver2008}. We perform final tests on the full-sized domain but determine the scaling factor $f$ on the domain down-sampled by a factor of 0.5 (this is done to speed up computations and to allow the possibility that currents have structure finer than the grid size, which is likely to be the case for real data). 

\begin{figure}[!hc]
  \begin{center}
   \includegraphics{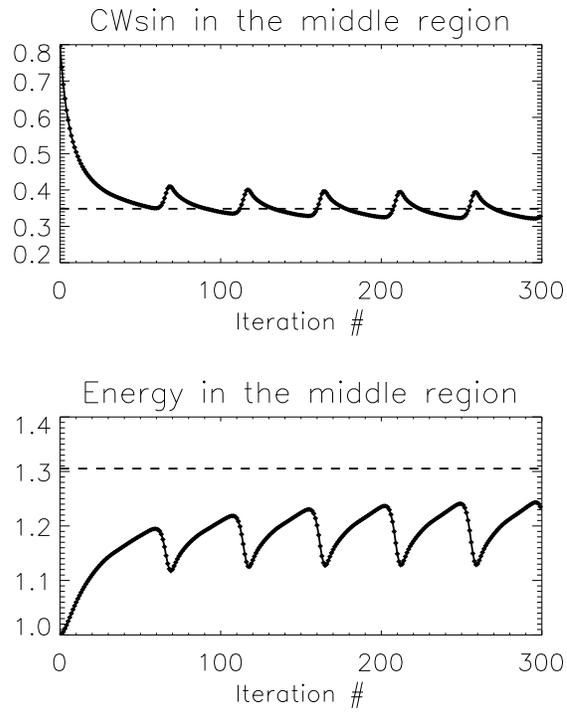} 
  \end{center}
 \caption{Values of CWsin and $E/E\pot$ in the center of the domain (the same region as used in \citet{Schrijver2008}) for the QGR calculation for one of the solar-like fields demonstrating oscillatory behavior. Different stages of this cycle are discussed in the text. The values of $E_\mathrm{ref}$ and CWsin$_{\mathrm{ref}}$ are shown as dashed lines.}  
  \label{alpha_oscill_fig1}
 \end{figure}

For both sets of loops (volume constraints for the down-sampled domain were only imposed at a small fraction of pixels, at about 1.7\% of the current-carrying volume and the bigger one covered about 6.6\% of the current-carrying volume) the solution for the pre-flare configuration does not converge with the II.b inputs. Instead it enters a remarkably stable oscillatory cycle with a period of $\approx50$ iterations. This cycle develops at $\approx200$ iterations, as shown in Figure~\ref{alpha_oscill_fig1}. We ran the code for a few thousands iterations to verify that the cycle is indeed stable. As energy slowly increases, a sheared arcade forms similar to the one in $\br$; but as the energy reaches its maximum and becomes most similar to $\br$, the field experiences drastic changes. Some of the current-carrying field lines rapidly ``escape'' the domain via the $y=-100$ boundary, what changes the \als values on these field lines (\als is set to zero), as explained in Section~\ref{sec_method}. In the stage of the cycle with the lowest energy most of the field lines from the core of the region connect to the $y=-100$ boundary and so carry no currents. This may be a valid force-free solution, though it is not consistent with the volume constraints which require \als to be non-zero in some points in the volume. When the \als values from the volume constraints are reimposed again at each iteration, the currents gradually build up again and the cycle repeats. The escape of field lines does not represent a physical evolution of the field as the iterations are not related to any physically meaningful time-like variable. The same oscillatory behavior is found in numerous experiments with this particular test-case. \citet{Schrijver2008} report that the pre-flare solution which we use as $\br$ did not fully converge either; it kept oscillating.

Below we discuss factors possibly causing the oscillations and a way to damp them. These factors are: (1) numerical noise in \als and therefore in the electric currents that appear even in current-free areas and (2) deviations of the input data from a force-free field (as we discuss below, $\br$ in this case is not exactly force-free even at full resolution). The damping that we consider allows the calculated field to have small variations in \als along field lines as well as small deviations of the solution from the volume constraints.

\clearpage

\begin{figure}[!hc]
 \begin{center}
   \includegraphics{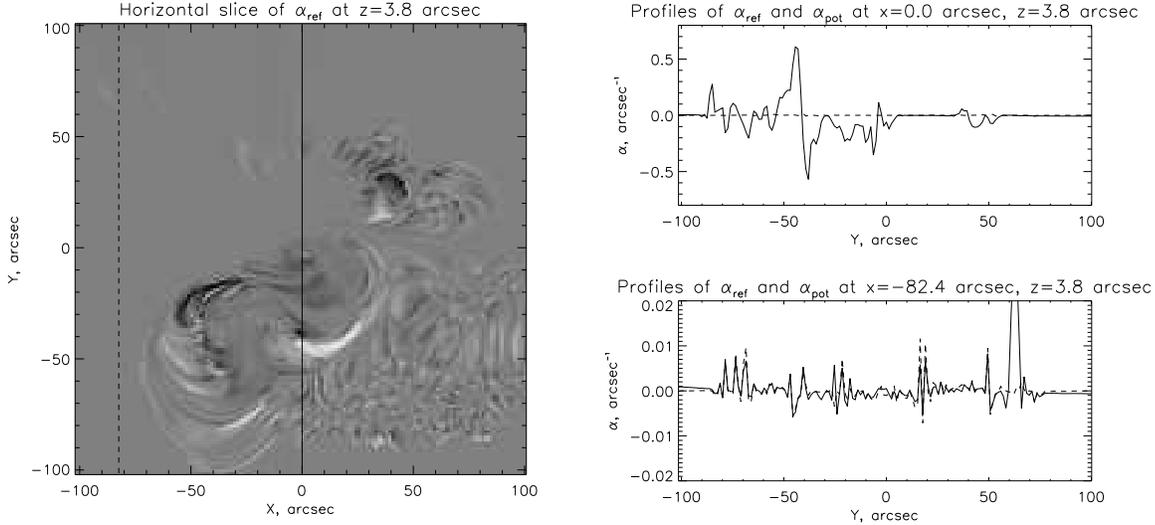} 
  \end{center}
 \caption{Left panel: image of horizontal slice of $\alpha_{\mathrm{ref}}$ in the pre-flare Wh$^{+}_{\mathrm{pp}}$ close to the lower boundary (the grayscale goes from $-0.8$ to $0.8$ arcsec$^{-1}$) and two profiles of $\alpha_{\mathrm{ref}}$ (solid line on both profiles) and $\alpha_{\mathrm{pot}}$ (dashed line on both profiles) in this slice. Top right panel: areas with significant currents. Bottom right panel: areas with no currents in both $\br$ and $\bp$. Variations in \als due to numerical uncertainties are $|\alpha|\lessapprox 0.005$ arcsec$^{-1}$. As we use the same numeric solver, the noise in our case is expected to be of the similar nature and magnitude.}
 \label{alpha_noise}
 \end{figure}

\begin{figure}[!hc]
  \begin{center}
   \includegraphics[width=8cm]{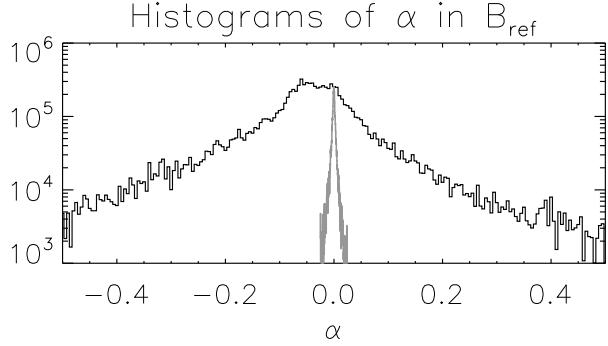} 
   \end{center}
  \caption{Histograms of $\alpha_{\mathrm{ref}}$ evaluated numerically on closed field (black line) and open field regions (gray line) in the pre-flare Wh$^{+}_{\mathrm{pp}}$. No currents are allowed on the open field by the GR scheme, used to calculate $\br$. Hence, the gray line shows the numerical noise. The distribution of the noise, evaluated from this plot, has a half width at half maximum of $\approx10^{-3}$ arcsec$^{-1}$.}
 \label{alpha_noise1}
 \end{figure}

The first factor is the influence of numerical noise when solving $\nabla\times\bvec^{(n+1)}=\alpha^{(n)}\bvec^{(n)}$ (Step~2 in the algorithm in Section~\ref{sec_method}), especially around sharp edges in $\alpha^{(n)}$, and with the artifacts introduced the by Fourier transforms around these edges. These effects introduce noise in the $\alpha^{(n+1)}$ values obtained in the next step. Figures~\ref{alpha_noise} and~\ref{alpha_noise1} clarify the amount of such noise and its relative size to the signal. In areas of closed field $|\alpha_{\mathrm{ref}}|\lessapprox 0.8\mbox{ arcsec}^{-1}$   and in the areas with open field (and hence no currents) $|\alpha_{\mathrm{ref}}|\lessapprox 5\times 10^{-3}\mbox{ arcsec}^{-1}$. The flux-weighted distribution of \als evaluated numerically in the current-free region has half width at half maximum of $\approx10^{-3}$ arcsec$^{-1}$. 

\begin{figure}[!hc]
 \begin{center}
   \includegraphics[width=12cm]{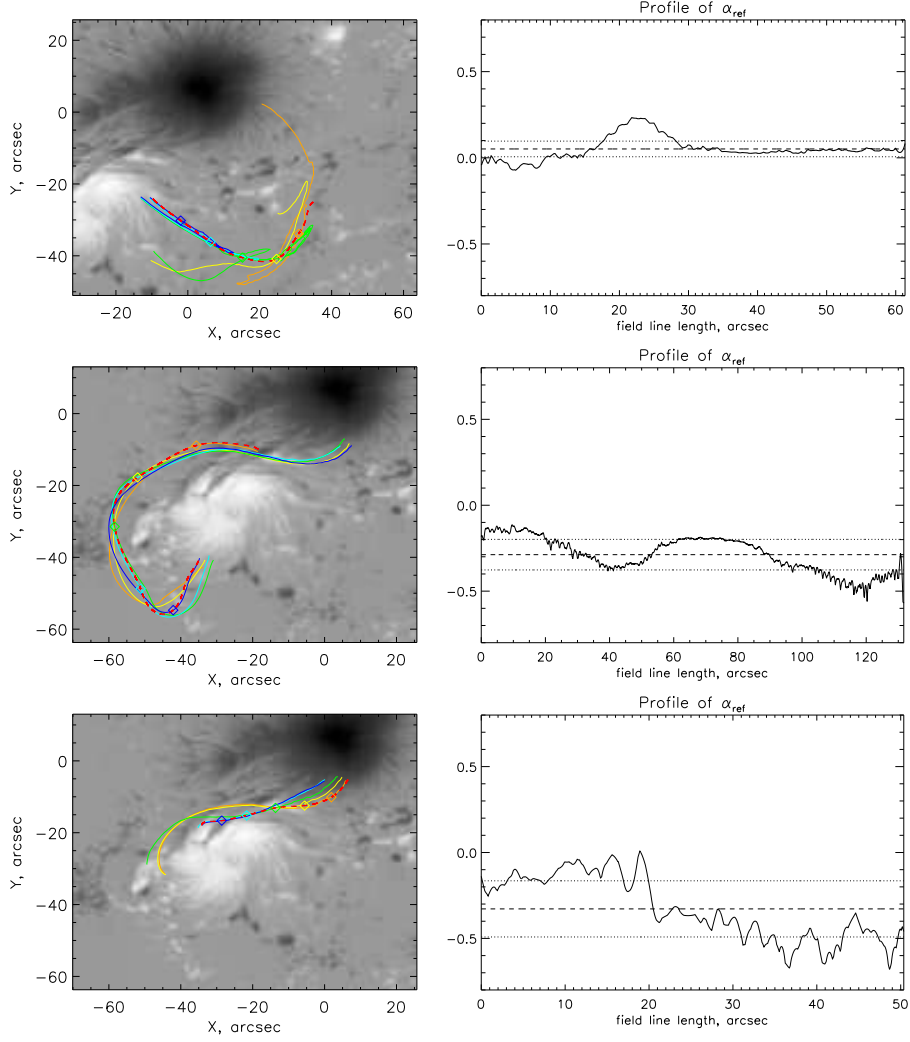} 
  \end{center}
 \caption{The variation in \als in the pre-flare Wh$^{+}_{\mathrm{pp}}$ field used as a solar-like test case. Left column: selected field lines of $\br$ (dashed red) and stream lines of $\jvec_{\mathrm{ref}}$ initiated at points along these field lines. Right column: profiles of $\alpha_{\mathrm{ref}}$ along these field lines. These panels indicate both small-scale and large-scale variation in $\alpha_{\mathrm{ref}}$ significantly above the noise threshold. The values of $\langle\alpha_{\mathrm{ref}}\rangle$ and $\langle\alpha_{\mathrm{ref}}\rangle \pm \sigma$ are shown as dashed and dotted lines respectively.}
 \label{alpha_constant}
 \end{figure}

The second factor is errors in the volume constraint data. In the case discussed in this section, $\br$ has significant non-zero magnetic forces from the perspective of QGR. As discussed in Section~\ref{sec_method}, for convergence QGR requires not only that the Lorentz force is small everywhere in the volume, but also that that the \textit{integrals} of the Lorentz force along the field lines are small. These conditions are not met for $\br$: as shown in Figure~\ref{alpha_constant}, \als changes substantially along central field lines. This is due to the Wh$_{\mathrm{pp}}^{+}$ solutions themselves not converging precisely during the GR iteration used to calculate them \citep[as mentioned in][]{Schrijver2008}.

To account for these issues we introduce two uncertainty thresholds: $\Delta\alpha_{\mathrm{err}}$ and $\Delta\alpha_{\mathrm{noise}}$. The first allows the solution to have \als values slightly different from the \als values imposed along loop trajectories and the second allows small variations of \als along field lines to damp numerical noise. A revised algorithm is formulated as follows, with the modifications relative to Section~\ref{sec_method} in bold.

\begin{enumerate}
	\item{Impose the volume constraints by setting $\alpha^{(n-1)}=\alpha_i$ along loop trajectories $\path_i$, \textbf{but only at points satisfying ${|\alpha^{(n-1)}-\alpha_i|\geq\Delta\alpha_{\mathrm{err}}}$}.}
	\item{Calculate updated field values $\bvec^{(n)}$ from Equation~\ref{curl_iter} subject to the prescribed boundary conditions. This equation is solved using a vector potential $\avec^{(n)}$ such that $\bvec^{(n)}=\nabla\times\avec^{(n)}$, so the divergence-free condition is satisfied to truncation error.}
  \item{Calculate an updated set of values for the force-free parameter $\alpha^{(n)}$: for every point in $\vol$, assign $\alpha^{(n)}=\langle\alpha^{(n-1)}\rangle$ \textit{averaged along the field line in $\bvec^{(n)}$ that passes through that point}, \textbf{but only at points satisfying  ${|\alpha^{(n)}-\langle\alpha^{(n-1)}\rangle|\geq\Delta\alpha_{\mathrm{noise}}}$. Otherwise retain the value of \als from the previous iteration}. If a field line leaves the domain through any boundary but the lower one, the value of $\alpha$ is set to zero along it (in common with the Wheatland~2007  GR scheme). This ensures that no currents go off to infinity so that the fields' energy remains finite.}
	\item{Repeat 1.-3. until $\bvec^{(n)}\approx\bvec^{(n-1)}$ and $\alpha^{(n)}\approx\alpha^{(n-1)}$ to within a tolerance and therefore $\nabla\times\bvec^{(n)}\approx\alpha^{(n)}\bvec^{(n)}$.}
\end{enumerate}

\begin{figure}[!hc]
  \begin{center}
   \includegraphics{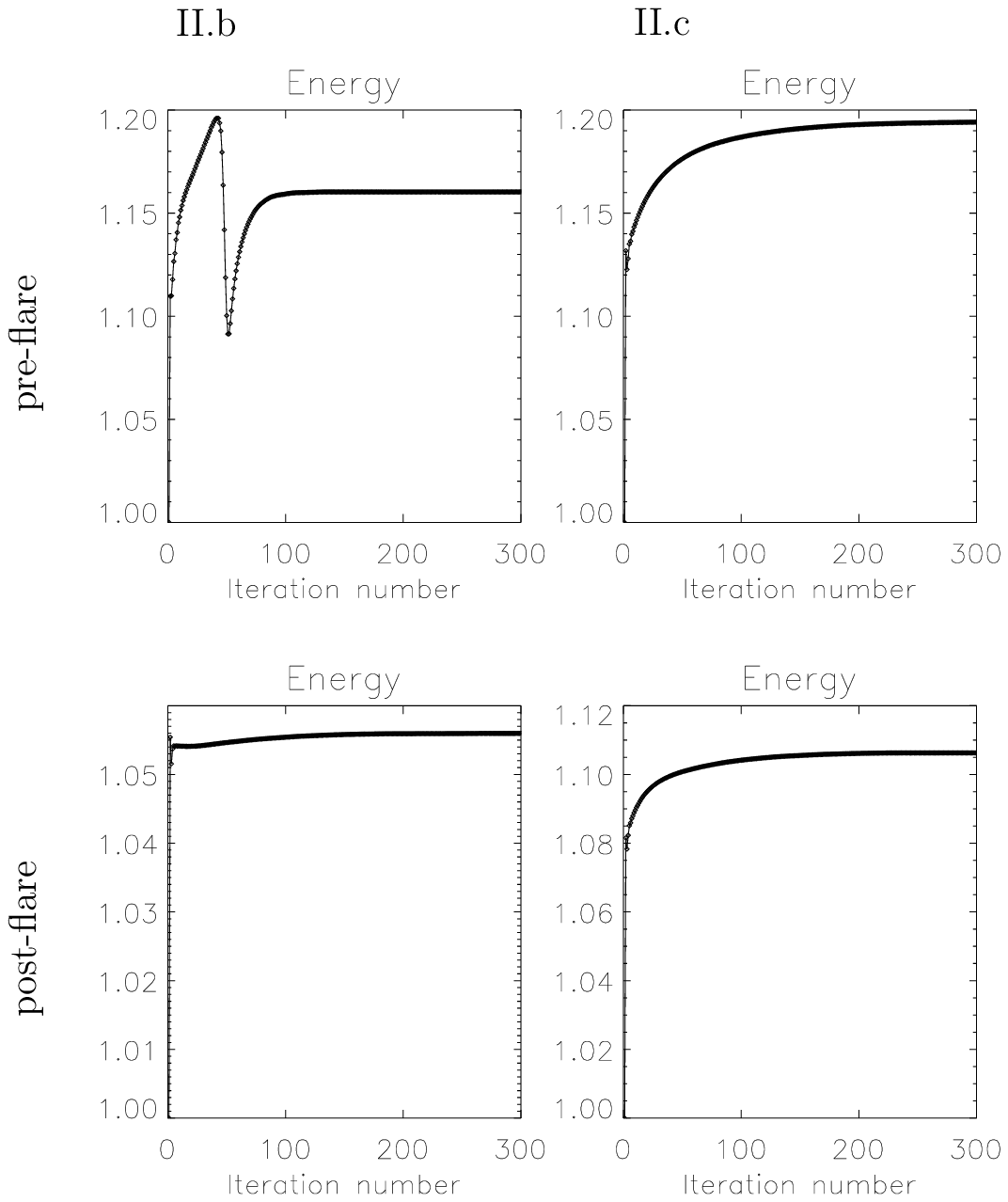} 
  \end{center}
  \caption{Energy $E/E_P$ at each iteration as demonstration of convergence of different schemes for the preflare and postflare Wh$^{+}_{pp}$ field fields. These particular plots correspond to the downsampled datacube for the case with fewer loops.}
 \label{wh_pp/energies}
 \end{figure}

For the case of the Wh$^{+}_{\mathrm{pp}}$ field we choose $\Delta\alpha_{\mathrm{err}}=\Delta\alpha_{\mathrm{noise}}\approx 5\times 10^{-4}$ arcsec$^{-1}$. For comparison, significant \als for the pre-flare $\br$ are $|\alpha_{\mathrm{ref}}|\propto0.8$ arcsec$^{-1}$, as shown in Figure~\ref{alpha_noise}. Two exceptions are the II.b pre-flare solution on the downsampled domain for fewer loops case and II.c pre-flare solution on the full-size domain for more loops case, for which the error threshold is increased to $5\times 10^{-3}$ arcsec$^{-1}$ in order to damp oscillations. Figure~\ref{wh_pp/energies} shows convergence plots for the II.b and II.c cases for the case with the fewer loops.

\begin{figure}[!hc]
 \begin{center}
   \includegraphics{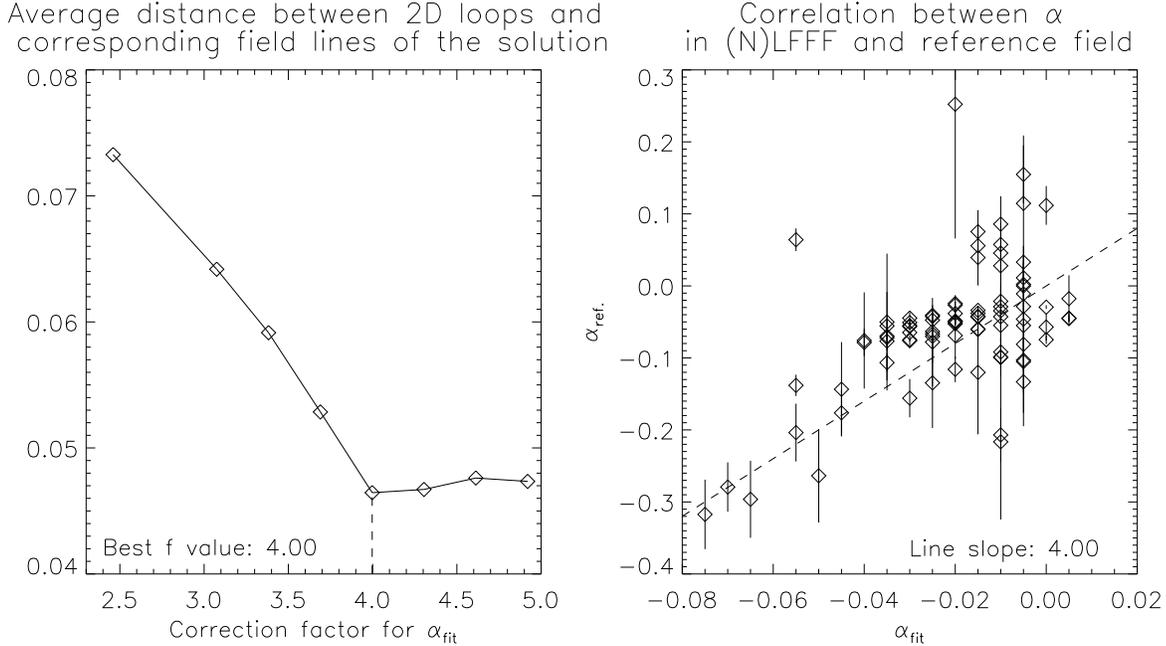} 
  \end{center}
 \caption{Left panel: average distance between loops from the pre-flare Wh$^{+}_{\mathrm{pp}}$ field and corresponding lines of $\bvec$ for the QGR solution in the same manner as in Figure~\ref{fact_llf}. The factor $f=4.0$, which yields the best-matching solution, is again close to scaling coefficient between $\alpha_{\mathrm{fit}}$ from MLM09 underestimates $\alpha_{\mathrm{ref}}$. Right panel: a scatter plot of $\alpha_{\mathrm{ref}}$ and $\alpha_{\mathrm{fit}}$ for individual loops and a line with the slope which equals to the best-matching $f$. Vertical error bars indicate the average variation of \als along loops in the reference field.}
 \label{fact_wh_pp}
 \end{figure}

\begin{figure}[!hc]
 \begin{center}
   \includegraphics{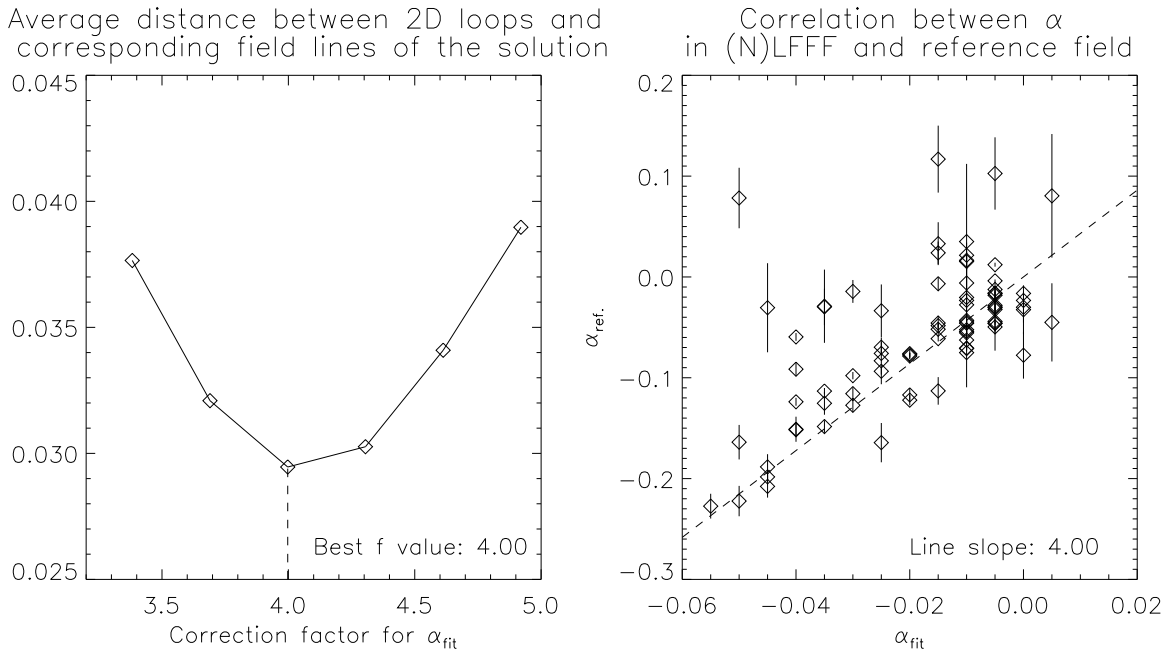} 
  \end{center}
 \caption{Same as Figure~\ref{fact_wh_pp}, but for the QGR solution for the post-flare reference field. The best-matching scaling factor $f$ is found to be the same as for the pre-flare field.}
 \label{fact_wh_pp_pf}
 \end{figure}

The correction factors $f$ are determined in the same way as in the Section~\ref{sec_llf}. We find a best-fitter $f=4.0$ for both pre- and post-flare data. In both cases this factor matches the coefficient between $\alpha_{\mathrm{ref}}$ and $\alpha_{\mathrm{fit}}$ (see Figures~\ref{fact_wh_pp} and~\ref{fact_wh_pp_pf}) for individual loops, which provides further evidence in favor of this method of estimating $f$.

\begin{figure}[!hc]
  \begin{center}
   \includegraphics{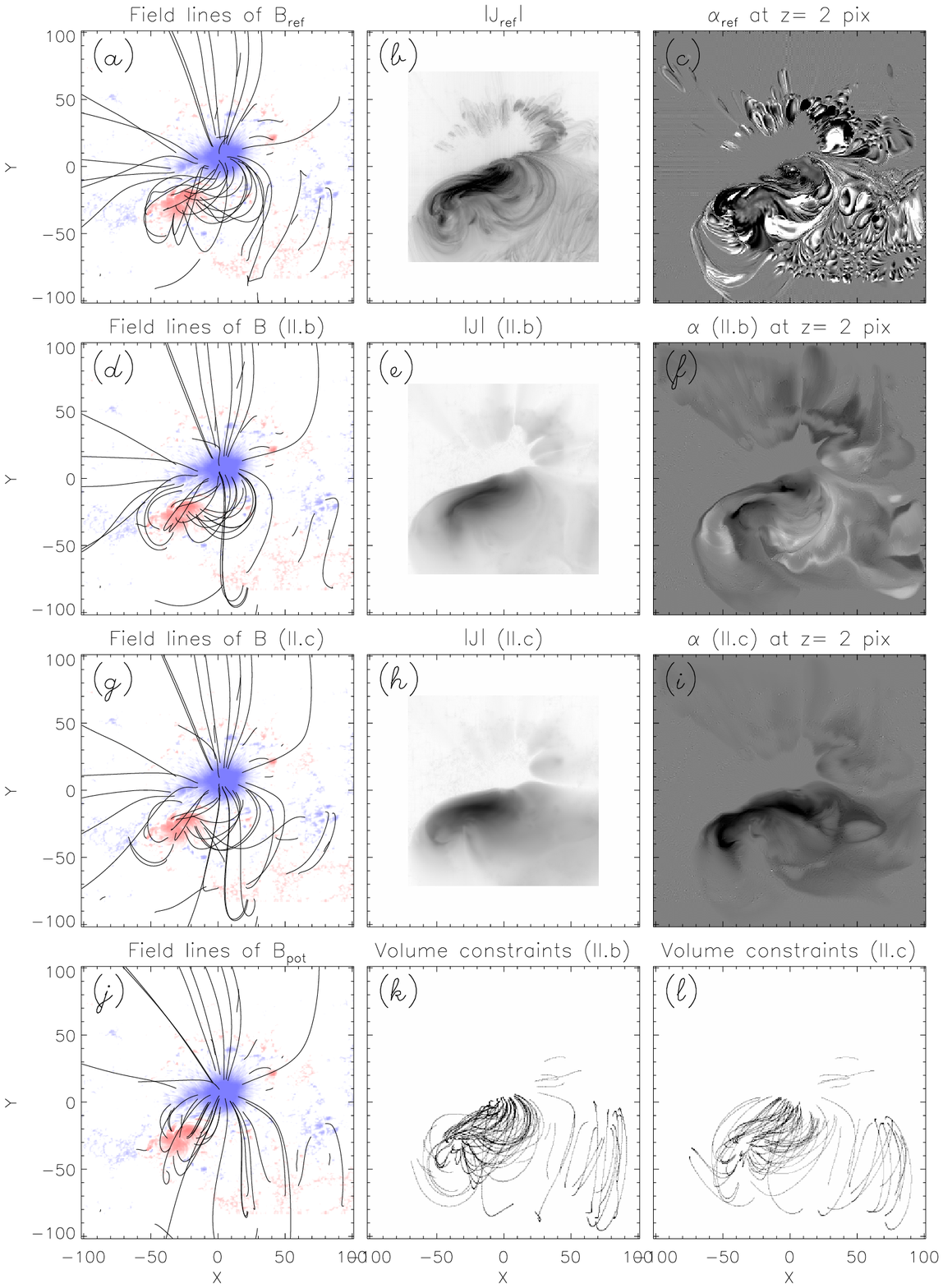} 
  \end{center}
  \caption{QGR solutions for the pre-flare Wh$_{\mathrm{pp}}^+$ reference field in full resolution for fewer loops case using schemes II.b and II.c (QGR with ideal and realistic loop input, that is, reconstructed from 2D loop projections --- refer to Figure~\ref{input_data_types}). Panels \textit{(a)}-\textit{(i)}: field lines, line-of-sight integrated magnitude of current and horizontal slice of \als for $\br$ and $\bvec$ for II.b and II.c. Panel \textit{(j)}: field lines of $\bp$ (all field lines are traced from the same starting points). Panels \textit{(h)}, \textit{(l)}: line-of-sight integrated volume constraints for II.b and II.c.}
 \label{wh_pp/fullres_1iic_25}
 \end{figure}

\begin{figure}[!hc]
  \begin{center}
   \includegraphics{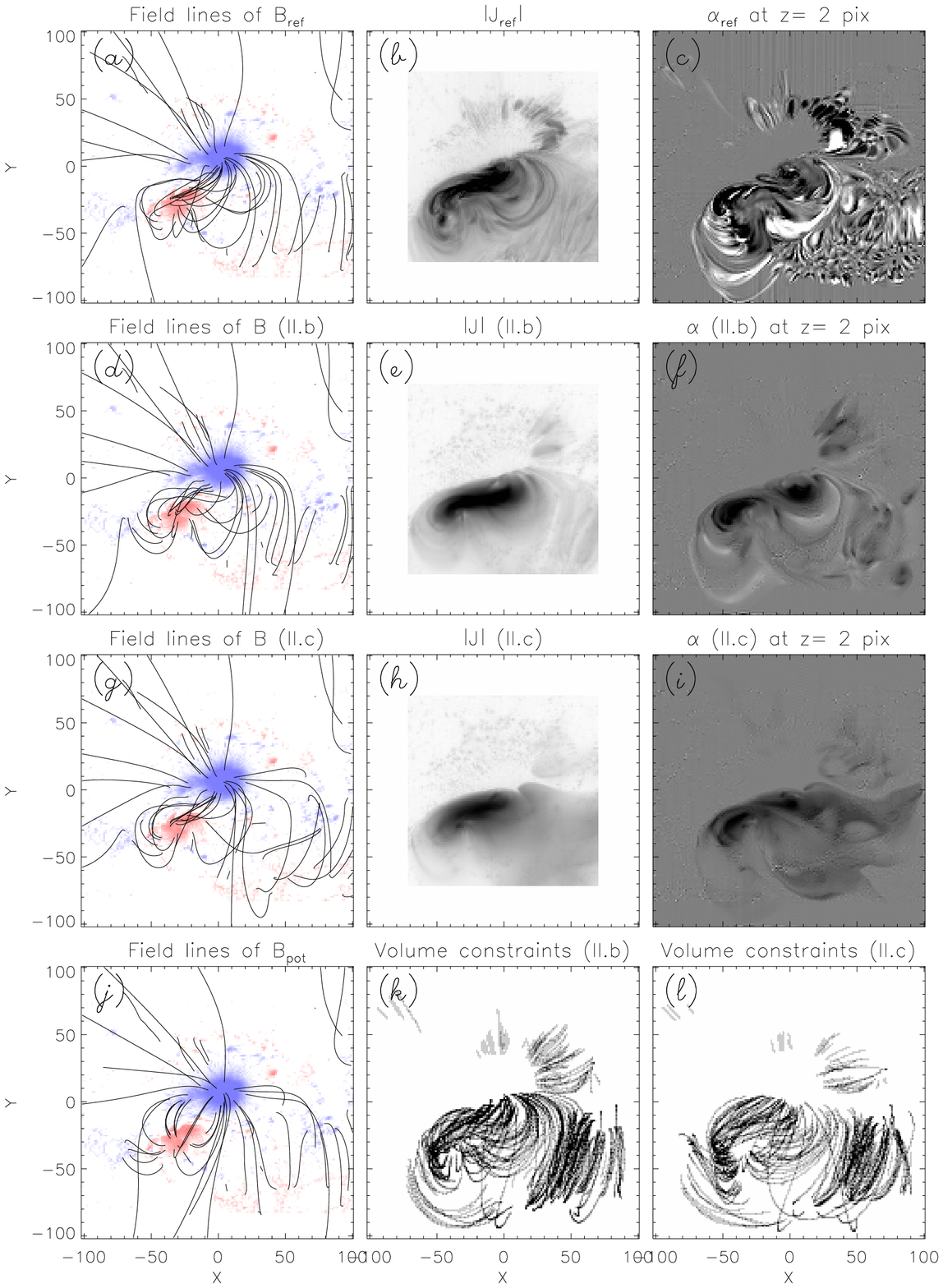} 
  \end{center}
  \caption{QGR solutions for the preflare Wh$_{\mathrm{pp}}^+$ reference field in half resolution for more loops using schemes II.b and II.c (QGR with ideal and realistic loop input, that is, reconstructed from 2D loop projections --- refer to Figure~\ref{input_data_types}). Panels \textit{(a)}-\textit{(i)}: field lines, line-of-sight integrated magnitude of current and horizontal slice of \als for $\br$ and $\bvec$ for II.b and II.c. Panel \textit{(j)}: field lines of $\bp$ (all field lines are traced from the same starting points). Panels \textit{(h)}, \textit{(l)}: line-of-sight integrated volume constraints for II.b and II.c.}
 \label{wh_pp/1iic_25_ml}
 \end{figure}

\begin{figure}[!hc]
  \begin{center}
   \includegraphics{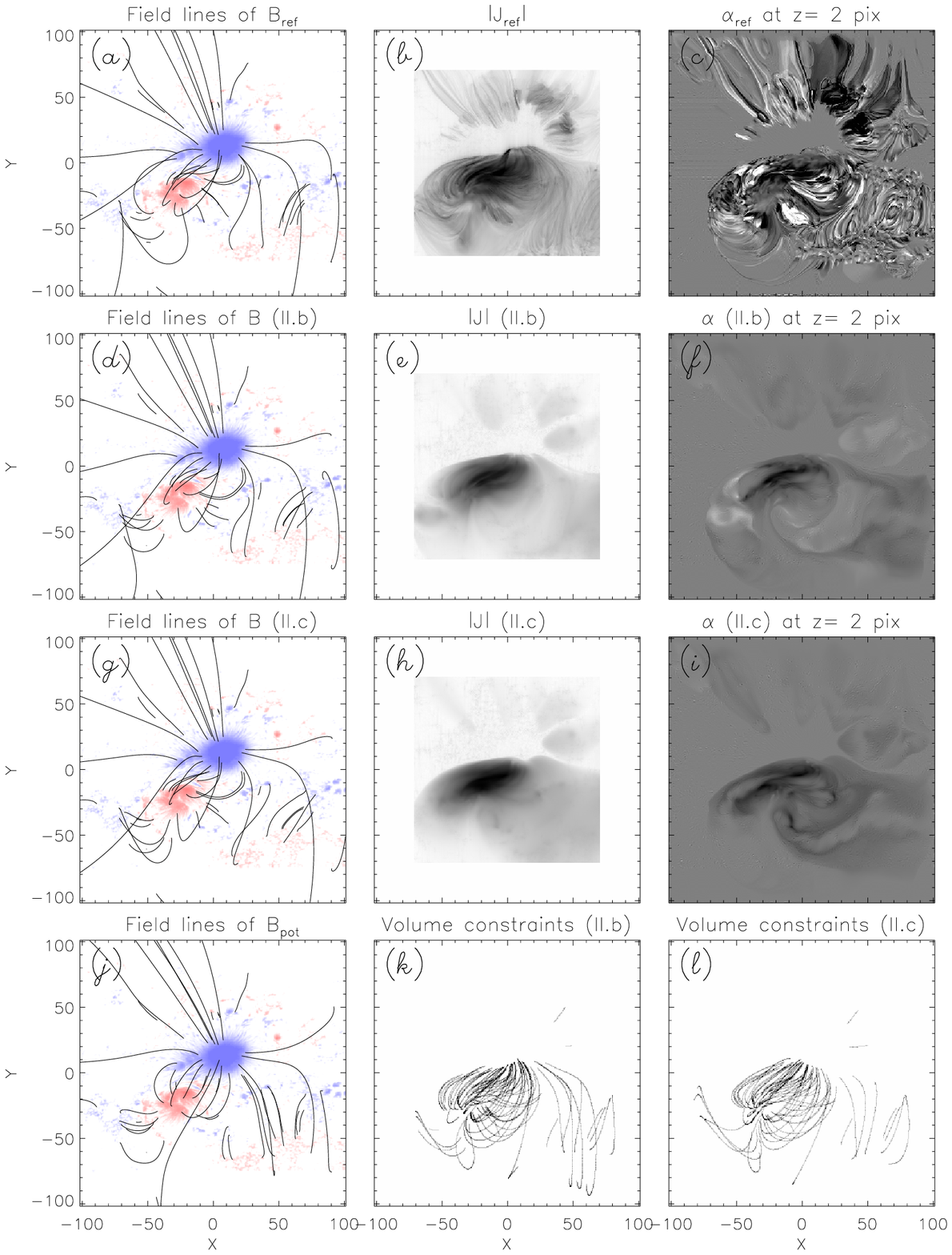} 
  \end{center}
  \caption{QGR solutions for the postflare Wh$_{\mathrm{pp}}^+$ reference field in full resolution using schemes II.b and II.c (QGR with ideal and realistic loop input, that is, reconstructed from 2D loop projections --- refer to Figure~\ref{input_data_types}). Panels \textit{(a)}-\textit{(i)}: field lines, line-of-sight integrated magnitude of current and horizontal slice of \als for $\br$ and $\bvec$ for II.b and II.c. Panel \textit{(j)}: field lines of $\bp$ (all field lines are traced from the same starting points). Panels \textit{(h)}, \textit{(l)}: line-of-sight integrated volume constraints for II.b and II.c.}
 \label{wh_postflare/fullres_1iic_25}
 \end{figure}

The results for the pre- and post-flare reference fields are summarized in Tables~\ref{table_wh_pp}-\ref{table_wh_pp_post-flare} and Figures~\ref{wh_pp/fullres_1iic_25}-\ref{wh_postflare/fullres_1iic_25}. In each case, the QGR reproduces the overall shape of field lines and the large-scale features of the current distribution. The reconstructed fields have $\ge50$\% of the free energy and $\ge25$\% of the relative helicity of the reference fields. The reconstructions using half resolution data are slightly inferior to those using full resolution data, based on the metrics in Tables~\ref{table_wh_pp}-\ref{table_wh_pp_post-flare}, but the solutions still reproduce at least half of the free energy and the quarter of helicity of the reference field and the large-scale structure of currents. 

\vspace{-1cm}\small{\begin{table}[!hc]
\begin{tabular}{m{1.0cm}m{1.5cm}m{1.5cm}m{1.5cm}m{1.5cm}m{1.5cm}m{1.5cm}c}
& & & & & & & \\
 & $\mbox{C}_{\mathrm{vec}}$ & $\mbox{C}_{\mathrm{CS}}$ & $1-E_n$ & $1-E_m$ & CWsin & $E/E\pot$ & $H(\bvec|\bp)$ \\
\hline
& & & & & & & \\
\multicolumn{8}{l}{\textbf{Half resolution}} \\
& & & & & & & \\
\multicolumn{8}{l}{\textit{Reference field}}\\
& 1.00 & 1.00 & 1.00 & 1.00 & 0.35 & 1.31 & 1.00 \\
& & & & & & & \\
\multicolumn{8}{l}{\textit{QGR with loop trajectories alone, fewer loops case}} \\
II.b & 0.98	& 0.98 & 0.83	& 0.84 & 0.37	& 1.16 & 0.62 \\
II.c & 0.97 & 0.97 & 0.79 & 0.80 & 0.32 & 1.20 & 0.36 \\
& & & & & & & \\
\multicolumn{8}{l}{\textit{QGR with loop trajectories alone, more loops case}} \\
II.b & 0.98	& 0.98 & 0.82	& 0.83 & 0.34	& 1.23 & 0.60 \\
II.c & 0.97	& 0.97 & 0.77	& 0.76 & 0.30	& 1.27 & 0.21 \\
& & & & & & & \\
\multicolumn{8}{l}{\textit{Potential field}}\\
 & 0.86 & 0.94 & 0.62 & 0.70 & --- &	1.00 & 0.00 \\
& & & & & & & \\ \hline
& & & & & & & \\
\multicolumn{8}{l}{\textbf{Full resolution}} \\
& & & & & & & \\
\multicolumn{8}{l}{\textit{Reference field}}\\
 & 1.00 & 1.00 & 1.00 & 1.00 & 0.24 & 1.32 & 1.00 \\
& & & & & & & \\
\multicolumn{8}{l}{\textit{QGR with loop trajectories alone, fewer loops case}} \\
II.b & 0.98 & 0.99 & 0.85 & 0.86 & 0.11 & 1.18 & 0.64 \\
II.c & 0.98 & 0.98 & 0.80 & 0.81 & 0.07 & 1.27 & 0.43 \\
& & & & & & & \\
\multicolumn{8}{l}{\textit{QGR with loop trajectories alone, more loops case}} \\
II.b & 0.98 & 0.99 & 0.83 & 0.84 & 0.09 & 1.26 & 0.62 \\
II.c & 0.97 & 0.97 & 0.77 & 0.77 & 0.08 & 1.30 & 0.29 \\
& & & & & & & \\
\multicolumn{8}{l}{\textit{Potential field}}\\
 & 0.86 & 0.94 & 0.62 & 0.70 & --- &	1.00 & 0.00 \\
\hline
\end{tabular}
\caption{\small{Metrics for the pre-flare reference field. The numbers for II.c solution are reported for the $f=4.0$. The downsampled fewer loops II.b case and full resolution more loops II.c case are unstable for $\Delta\alpha_{\mathrm{err}}=5\times 10^{-4}$ arcsec$^{-1}$; the reported values in these cases are for $\Delta\alpha_{\mathrm{err}}=5\times 10^{-3}$ arcsec$^{-1}$.}}
\label{table_wh_pp}
\end{table}}

 \voffset=-1.5cm
 \hoffset=-1cm

 \voffset=-1.5cm
 \hoffset=-1cm

 \begin{table}
	\centering
\begin{tabular}{m{1.0cm}m{1.9cm}m{1.9cm}m{1.9cm}m{1.9cm}m{1.9cm}m{1.9cm}c}
& & & & & & & \\
 & $\mbox{C}_{\mathrm{vec}}$ & $\mbox{C}_{\mathrm{CS}}$ & $1-E_n$ & $1-E_m$ & CWsin & $E/E\pot$ & $H(\bvec|\bp)$ \\
& & & & & & & \\ \hline
& & & & & & & \\
\multicolumn{8}{l}{\textbf{Half resolution}} \\
& & & & & & & \\
\multicolumn{8}{l}{\textit{Reference field}}\\
 & 1.00 & 1.00 & 1.00 & 1.00 & 0.13 & 1.16 & 1.00 \\
& & & & & & & \\
\multicolumn{8}{l}{\textit{QGR with loop trajectories alone}}\\
II.b & 0.99	& 0.99 & 0.86	& 0.87 & 0.10	& 1.07 & 0.48 \\
II.c & 0.99 & 0.99 & 0.88 & 0.87 & 0.07 & 1.13 & 0.63 \\
& & & & & & & \\
\multicolumn{8}{l}{\textit{Potential field}}\\
 & 0.94 & 0.97 & 0.76 & 0.80 & --- &	1.00 & 0.00 \\
& & & & & & & \\ \hline
& & & & & & & \\
\multicolumn{8}{l}{\textbf{Full resolution}} \\
& & & & & & & \\
\multicolumn{8}{l}{\textit{Reference field}}\\
 & 1.00 & 1.00 & 1.00 & 1.00 & 0.17 & 1.14 & 1.00 \\
& & & & & & & \\
\multicolumn{8}{l}{\textit{QGR with loop trajectories alone}}\\
II.b & 0.99 & 0.99 & 0.89 & 0.88 & 0.13 & 1.09 & 0.50 \\
II.c & 0.99 & 0.99 & 0.89 & 0.88 & 0.10 & 1.14 & 0.69 \\
& & & & & & & \\
\multicolumn{8}{l}{\textit{Potential field}}\\
 & 0.93 & 0.97 & 0.75 & 0.80 & --- &	1.00 & 0.00 \\
& & & & & & & \\ \hline
\end{tabular}
\caption{Metrics for the QGR results for the post-flare reference field. The numbers for II.c are reported for the $f=4.0$ solution. For notation, refer to Table~\ref{input_data_types}.}
\label{table_wh_pp_post-flare}
\end{table}

 \voffset=-1.5cm
 \hoffset=-1cm
 
 \voffset=-1.5cm
 \hoffset=-1cm

\section{Discussion and Conclusions}\label{sec_summ}

In this study we demonstrate that coronal loops provide a useful source of information for determining the structure of the coronal magnetic field. While the observed loops do not cover all of the coronal volume, they provide information about the \textit{shape of the coronal field lines}, which boundary data alone lack. 

We demonstrate a method that constructs nonlinear force-free fields using line-of-sight magnetograms and coronal loops observed in the plane-of-sky projection. This may mitigate the problems NLFFF schemes encounter with currents determined from vector magnetograms \citep{Demoulin1997b}. The loops are first approximated by lines of constant-\als fields, with different \als values for each loop. This is done using an existing scheme developed by \citet{Malanushenko2009b} which we refer to as the MLM09 fit in this paper. The approximate \als values along the approximate loop trajectories are treated as volume constraints in a quasi Grad-Rubin algorithm, using the code modified from \citet{Wheatland2009}. 

The method, which we refer to as the Quasi Grad-Rubin method (or QGR) is tested on several nonlinear force-free fields and the results demonstrate good performance of the method. While traditional extrapolations of coronal magnetic fields have been found to provide poor matches to coronal features observed in X-rays and EUV \citep{DeRosa2009}, the fields created by QGR are \textit{constructed} with the effort to match observed coronal features and thus may provide a more realistic model of the actual coronal magnetic field. 

The problem of constructing a nonlinear force-free field is typically viewed as a boundary value problem requiring an extrapolation of the field from the boundaries to the volume of the corona. However, thorough this paper we purposely avoid referring to QGR as an extrapolation scheme, because it is not. It is a mixture of extrapolation of magnetic field and \textit{inter}polation of electric currents. The reason we tend to view the step of filling the volume with \als values as an interpolation-like procedure is as follows. At each iteration, \als is averaged along lines of the field at the present iteration. If the solution has not converged yet, this field is different from the one on the previous iteration, so \als is averaged across lines of the field of the previous iteration (see Section~\ref{sec_method}). Substantial differences in \als values on such field lines result in extreme values getting ``spread'' through the volume and the values of \als being ``smoothed'' along field lines. Smaller differences between the fields from two consecutive iterations should result in \als smoothing out on shorter distances. So on each consecutive iteration $n$ the \als values are smoothed across field lines to a distance which depends on the magnitude of the angle between $\bvec^{(n)}$ and $\bvec^{(n-1)}||\jvec^{(n)}$, and therefore on the Lorentz force at the $n$-th iteration. The process therefore results in a smooth distribution of \als in the volume and decreasing the Lorentz forces implies smaller-scale changes in this already smooth distribution. The described scheme cannot produce \als bigger in magnitude than the volume constraints and it tends to produce a smooth transition of \als between these constraints in such a way as to minimize Lorentz forces. This explains the interpolation-like nature of QGR with respect to \al. This scheme does not resolve fine structure of the ``interpolated'' variable (in this case, \al), as no interpolation scheme can, but it successfully approximates general trends, as expected from an interpolation scheme. 

We also develop a way to deal with the uncertainties in the input data and the numeric noise. The uncertainties in the observables produce inconsistency with a force-free solution. As such uncertainties are expected, this is an important feature of the method. QGR in the form described in Section~\ref{sec_karel} allows the volume constraints not to be re-imposed if the average \als on a given field line which passes through a given constraint point is within a small prescribed amount of $\alpha_i$ of the constraint. This means that a magnetic field which is force-free but imperfectly matches given volume constraints for \als would not be changed by the method. As any numerical scheme, QGR is also prone to numerical noise. We are able to determine the range of this noise for a given problem. The method assumes that \als below this noise level is numerically not distinguished from zero. It also assumes that average \als along a given field line may vary within this noise range. It therefore does not replace \als by the newly defined average along the field line if that average differs to less than the numerical noise threshold from the previously determined value. 

We noted that fitting loops with lines of \citet{Chiu1977} constant-\als fields results in the underestimation of \als and verify that it is at least partly due to a difference in the size of the volumes which contain currents (finite in reference cases and half space in the fields used for fitting, see Section~\ref{sec_dipole}). We determine that the underestimation coefficient is roughly the same for most loops and that this coefficient may be determined from observables (projected loops). Applying this determined coefficient leads to a good match between the reference field and the model, as demonstrated in Section~\ref{sec_tests}. A rigorous proof of the nature of such a coefficient and its analytic evaluation are subjects of future studies. 

We do not find substantial differences when reconstructing a reference field on a full size domain or on a down-sampled domain. This might be due to the smoothness of fields created by the QGR due to its interpolation-like nature for the reasons discussed above. This is an important result, as the test case in Section~\ref{sec_karel} has structure of currents finer than the grid size in the down-sampled test, which may also be the case when modeling coronal fields.  

While developed for currents approximated from loops in EUV and X-ray images, QGR yields better results when currents are measured exactly, e.g., from vector magnetograms (II.b inputs in Sections~\ref{sec_llf} and~\ref{sec_karel}). This gives hope that as vector magnetograms become more applicable for NLFFF modeling (at least in the cores of active regions), the performance on the II.b level could be achieved. This method could also benefit from the exact knowledge of the 3D shape of the loops, e.g., drawn from STEREO and SDO satellites combined. 

Overall we find that the method developed in this paper is able to recover over half of the free energy and over a quarter of the helicity for the solar-like test case fields, which is more than was reported for previously tested methods \citep{Metcalf2008, Schrijver2008, DeRosa2009}. The method recovers large-scale features of the field well, such as structure of currents, shape of field lines and the connectivity of the field, but it fails to resolve fine structure. We nonetheless find that the large structure determines at least half of the free energy and a quarter of the relative helicity and therefore QGR may be used to provide estimates of these quantities. 

This work was supported by AIA contract NNG04EA00C to the Lockheed Martin Advanced Technology Center through a grant to Montana State University, 
in collaboration with the University of Sydney. \textit{Hinode} is a Japanese mission developed and launched by ISAS/JAXA, with NAOJ as domestic partner and NASA and STFC (UK) as international partners. It is operated by these agencies in co-operation with ESA and NSC (Norway). 

\bibliography{c:/localtexmf/bib/short_abbrevs,c:/localtexmf/bib/full_lib,c:/localtexmf/bib/my_bib}
\end{document}